\documentclass[11pt]{article}
\pdfoutput=1

\usepackage{jheppub}

\usepackage{amssymb}
\usepackage{amsfonts}
\usepackage{graphicx}
\usepackage[dvipsnames,table]{xcolor}
\usepackage{epstopdf}
\usepackage{dcolumn}
\usepackage{amsmath}
\usepackage{latexsym,bm}
\usepackage{amsthm}
\usepackage{slashed}
\usepackage{float}
\usepackage{color}
\usepackage{url}
\usepackage{longtable}
\usepackage{subfig}
\usepackage{tikz}
\usepackage[all,cmtip]{xy}
\usepackage{rotating}

\usepackage{multirow}
\usepackage{longtable}

\makeatletter

\usepackage{tikz}
\usetikzlibrary{arrows}
\usetikzlibrary{arrows.meta}
\usetikzlibrary{shapes.geometric,calc,arrows, positioning,shapes.misc,decorations.markings}
\tikzset{
  big arrow/.style={
    decoration={markings,mark=at position 1 with {\arrow[scale=2,#1]{>}}},
    postaction={decorate},
    shorten >=0.4pt},
  big arrow/.default=black}
  
\pgfdeclarelayer{edgelayer}
\pgfdeclarelayer{nodelayer}
\pgfsetlayers{edgelayer,nodelayer,main} 
\tikzstyle{none}=[inner sep=0pt] 

\tikzstyle{NodeCross}=[draw, shape=circle, cross out, inner sep=0pt, minimum size=6pt,line width=0.25mm]
\tikzstyle{Circle}=[draw, shape=circle, black,  fill=black, inner sep=0pt, minimum size=6pt]
\tikzstyle{Star}=[draw, shape=star, fill=black, star points=8, inner sep=0pt, minimum size=8pt]

\tikzstyle{DashedLine}=[-, densely dashed, line width=0.25mm]
\tikzstyle{DottedLine}=[-, dotted, line width=0.25mm]
\tikzstyle{ThickLine}=[-, line width=0.25mm]
\tikzstyle{ArrowLineRight}=[-, -{Stealth[scale=1.75]}, line width=0.1mm, scale=5]
\tikzstyle{RedLine}=[-, draw={rgb,255: red,191; green,0; blue,0}, fill=none, line width=0.25mm]
\tikzstyle{DottedRed}=[-, dotted, draw={rgb,255: red,191; green,0; blue,0}, fill=none, line width=0.25mm]
\tikzstyle{DashedLineThin}=[-, densely dashed, line width=0.125mm, fill=none, draw=black]
\tikzstyle{ArrowLineRed}=[-, -{Stealth[scale=1.75]}, draw={rgb,255: red,191; green,0; blue,0}, line width=0.25mm, scale=5]

\newcommand{\be}{\begin{equation}}
\newcommand{\ee}{\end{equation}}
\newcommand{\ba}{\begin{aligned}}
\newcommand{\ea}{\end{aligned}}

\newcommand{\bea}{\begin{eqnarray}}
\newcommand{\eea}{\end{eqnarray}}

\def\unit{{1\kern-.65ex {\rm l}}}
\def\1{{1\kern-.65ex {\rm l}}}

\newcount\hour \newcount\minute
\hour=\time \divide \hour by 60
\minute=\time
\def\now{%
\ifnum \hour<13
  \ifnum \hour=0 \advance \hour by 12 \number\hour:\else \number\hour:\fi%
     \ifnum \minute<10 0\fi%
     \number\minute%
\ A.M.%
\else \advance \hour by -12 \number\hour:%
  \ifnum \minute<10 0\fi%
  \number\minute%
  \ P.M.%
\fi%
}

\makeatother

\def\mb{\mathbb}

\def\mc{\mathcal}
\def\bp{\begin{pmatrix}}
\def\ep{\end{pmatrix}}

\graphicspath{ {figs/} }

\title{Statistics of Base Polytopes in F-theory}

\preprint{\today \hspace*{0.1in} }

\author[1]{Washington Taylor}
\emailAdd{wati@mit.edu}
\author[2,3]{Yi-Nan Wang}
\emailAdd{ynwang@pku.edu.cn}
\author[2]{Yihang Yu}
\emailAdd{yyh1322026225@stu.pku.edu.cn}

\preprint{MIT-CTP/5924}

\affiliation[1]{Center for Theoretical Physics\protect\\
 Department of Physics\protect\\
 Massachusetts Institute of Technology\protect\\
 77 Massachusetts Avenue, Cambridge, MA 02139, USA}

\affiliation[2]{School of Physics, Peking University, \protect\\
Beijing 100871, China}

\affiliation[3]{Center for High Energy Physics, Peking University, \protect\\
Beijing 100871, China}

\abstract{We propose a new statistical ensemble of toric bases for elliptic Calabi-Yaus used in F-theory models, by focusing on only the convex hull of the base, i.e., the base polytope. This physically motivated coarse-graining greatly simplifies the combinatorial complexity of the part of the 4d F-theory landscape with toric bases. We develop a Monte Carlo approach that randomly samples the base polytopes within fixed boxes, with proper statistical weights. We first apply the algorithm to the set of 2d base polytopes, generating an enlarged set of toric 2d bases that include certain types of codimension-two (4,6) points, and we validate our approach against exact numbers. We then explore the set of 3d base polytopes which fit in a set of ``maximal'' 3d boxes, and estimate the total number of inequivalent 3d base polytopes to be $10^{85}$--$10^{90}$. We provide statistical data such as the distribution of non-Higgsable gauge groups on these bases. Amusingly, a similar method can also be applied to generate reflexive polytopes in various dimensions. In both the reflexive and base polytope cases, the number of relevant polytopes obeys a Gaussian distribution as a function of the number of vertices, which can be understood in terms of other results on random polytopes in the math literature.} 

\begin{document}

\maketitle

\section{Introduction}

F-theory~\cite{Vafa:1996xn,Morrison:1996na,Morrison:1996pp,Weigand:2018rez}
provides a vast geometric landscape of UV complete quantum gravity
theories in even spacetime
dimensions~\cite{Kumar:2009ac,Taylor:2011wt,Morrison:2012np,Morrison:2012js,Martini:2014iza,Taylor:2015isa,Halverson:2015jua,Taylor:2015ppa,Watari:2015ysa,Taylor:2015xtz,Halverson:2017ffz,Taylor:2017yqr,Cvetic:2019gnh,Tian:2020yex,Morrison:2023hqx}. To
get supergravity theory in $(10-2d)$-dimensional spacetime, we
consider F-theory on an elliptic Calabi-Yau $(d+1)$-fold $X_{d+1}$
over a compact complex $d$-dimensional base $B_d$, defined as 
\be
X_{d+1}:\ y^2=x^3+fxz^4+gz^6\,,
\label{eq:Weierstrass}
\ee
where $f\in\mc{O}(-4K_{B_d})$, $g\in\mc{O}(-6K_{B_d})$, $[x:y:z]$ are the coordinates of $\mb{P}^{2,3,1}$.

While generic F-theory compactifications are intrinsically
nonperturbative, and precise details of the low-energy theory are
difficult to compute, this approach gives a broader global picture of
large connected components of the
landscape of string theory compactifications than any other known
approach. 
The complete F-theory landscape in $(10-2d)$-dimensions can be built up with the following steps:
\begin{enumerate}
\item Classifying topologically distinct bases $B_d$. For the case of
  $d=2$, the set of smooth base surfaces without codimension-two (4,6)
  locus are classified for the toric~\cite{Morrison:2012js} and
  semi-toric subsets~\cite{Martini:2014iza}, and partially classified
  for the general non-toric cases~\cite{Taylor:2015isa}. For the case
  of base threefolds, which are relevant for 4d model building, there
  is not even a complete classification of all the allowed base
  threefolds in the toric subset, and very little is known about the
  full space of non-toric threefold bases. Nonetheless, there have been
  different approaches to probe the subset of smooth toric base
  threefolds~\cite{Halverson:2015jua,Taylor:2015ppa,Halverson:2017ffz,Taylor:2017yqr},
  including explicit constructions and Monte Carlo analysis. 

\item On a fixed base $B_d$, classifying different elliptic fibration
  structures leading to topologically distinct elliptic Calabi-Yau
  $(d+1)$-folds $X_{d+1}$. Physically, these distinct choices can give
  rise to different enhanced (tuned) geometric gauge groups, matter
  curves and geometric Yukawa couplings for different $d$. 
While the Weierstrass model (\ref{eq:Weierstrass}) can be directly
tuned to realize a variety of nonabelian gauge factors, matter
representations, and low-rank abelian gauge factors,
the full set of possible fibrations is
not completely understood;  in the toric context, there has  been some partial
exploration of the tuning possibilities for CY3 hypersurfaces in
reflexive polytopes~\cite{Huang:2018esr,Huang:2019pne,Abbasi-nt}. One can also potentially add frozen singularities (O7$_+$ branes)~\cite{Tachikawa:2015wka,Bhardwaj:2018jgp,Morrison:2023hqx,Oehlmann:2024cyn}.

\item For the $d\geq 3$ cases, after fixing the Weierstrass model, one needs to classify the discrete choices of $G_4$ flux. In particular, the elliptic CY4 with the currently known largest $h^{3,1}=303148$ gives rise to the largest number of 4d flux vacua in the currently known string theory literature, which is estimated to be $\sim 10^{224000}$ (with self-duality condition)~\cite{Taylor:2015xtz}.

\item After fixing the $G_4$ flux, one should in principle compute the scalar potential $V$ and stabilize the complex structure moduli, K\"{a}hler moduli and D3-brane moduli, for the $d=3$ cases. 

\end{enumerate}

The first step in developing a comprehensive global picture of the 4d
F-theory landscape is thus the classification of the set of 3d compact
complex bases that support an elliptic Calabi-Yau fourfold.  Unlike in
the case of 2d bases, this set of bases is not completely classified
(although it is known to be finite \cite{Di_Cerbo_2021}), and even the relatively
controlled domain of toric 3d bases is only partially understood.
In this paper, we revisit the problem of classifying toric base
threefolds $B_3$ for 4d F-theory. In fact, if one tries to classify
all the possible bases up to isomorphism, when $h^{1,1}(B_3)$ is
large, there is a huge number of different bases related to each other
via flips and flops. For example, the base $B_3$ of the elliptic CY4
with the largest $h^{1,1}=303148$ has at least $7.5\times10^{45766}$
flip and flop phases~\cite{Wang:2020gmi}.  Previous studies of the
set of toric threefold bases have
included this kind of redundancy.

To remove this redundancy, we will try a new and different approach, by identifying the base threefolds $B_3$ whose toric fan has the same convex hull. In other words, we are classifying the different base polytopes $\Delta_b$, up to a $GL(3,\mb{Z})$ transformation, while ignoring the different possible triangulations.

There are several motivations for considering this
triangulation-independent polytope ensemble.  Physically, many
features of the associated 4d F-theory models are independent of the
triangulation of the base; most notably, the rigid (geometrically
non-Higgsable) gauge group \cite{Morrison:2014lca} of the theory does
not depend on the specific choice of fan for the base polytope.  Additionally,  by giving a more coarse-grained picture of the toric 4d F-theory landscape, this ensemble may provide a clearer global picture of this set of string vacua.  Since we do not have any good way of determining the cosmological measure, it is important to investigate different ways of charting this landscape at different levels of detail.

Of course, a given base polytope $\Delta_b$ may not lead to a smooth
toric threefold. In other words, the base may possess terminal singularities for any triangulation. A
simple example of this case would be the base polytope with vertices
$\{(1,0,0)$, $(0,1,0)$, $(0,0,1)$, $(-1,-1,-2)\}$. In this paper we
would generally allow such kinds of base polytope to be a part of our
enlarged base ensemble.
Note that such singular bases 
are a somewhat
generic feature of the 3d bases expected in the F-theory landscape.

 In this paper we assume  some familiarity with toric geometry and the
 structure of toric threefold bases for F-theory; much of the
 technical apparatus used is described in more detail in
 \cite{Taylor:2015ppa,Taylor:2017yqr}, and we refer the reader to those references for
 more background.

We now describe
the precise combinatoric conditions for a valid 3d base polytope $\Delta_b$, while for general $d$-dimensional base polytopes the conditions are simply generalized by replacing the 3d vectors $u$, $v_i$ and $\vec{0}$ with $d$-dimensional vectors. Denoting the vertices of $\Delta_b$ by $\{v_i\}$, the conditions are:
\begin{enumerate}
\item{$v_i=(v_{i,1},v_{i,2},v_{i,3})$ are all primitive rays, such that gcd$(v_{i,1},v_{i,2},v_{i,3})=1$.}
\item{The origin $(0,0,0)$ is in the interior of $\Delta_b$, to ensure that the base threefold $B_3$ is compact.}
\item{Consider the $\mc{G}$-polytope, whose lattice points
 have a
  one-to-one correspondence to the monomial generators of $\mc{O}(-6K_B)$. This is defined as
\be
\mc{G}=\{u\in\mb{Z}^3|\forall v_i\ ,\ \langle u,v_i\rangle\geq -6\}\,.
\ee
 In order for $B_3$ to support an elliptic CY4 (which is at finite
 distance in the moduli space of all CY4s, but which may include some singularities), the condition is that the origin $(0,0,0)$ is in the interior of $\mc{G}$~\cite{Taylor:2017yqr}. 

}
\end{enumerate}

It turns out that the full, explicit classification of such base polytopes $\Delta_b$ is still impractical to achieve. In this paper, we will attack the combinatorics problem by a fully parallelizable Monte Carlo approach, to get an approximate picture of the whole landscape of base polytopes.

In practice, we will randomly generate (sample) base polytopes in a fixed 3d box $\mc{B}$ by choosing $n$ random points in the box, and give an appropriate weight factor to each sampling taking into account different statistical redundancies. We find that the dominating redundancy factor is given by the combinatoric factor
\be
R(l(\Delta_b),m,n)=\sum_{k=0}^{m}(-1)^{m-k}\binom{m}{k}(l(\Delta_b)-m)^n\sum_{p=m}^{n}\binom{n}{p}\left(\frac{k}{l(\Delta_b)-m}\right)^p\,,
\ee
where $l(\Delta_b)$ denotes the number of lattice points in the base polytope $\Delta_b$ given by the convex hull of the $n$ random points. $m$ is the number of vertices of $\Delta_b$. This formula can be generally applied to the random generation of polytopes in any dimension $d$.

Interestingly, if we treat the number of vertices $m$ as a random
variable, we find that the number of samplings with a given $m$ obeys
an approximate Gaussian distribution. As a consequence, the $\log$ of
the redundancy factor also obeys a Gaussian distribution to high accuracy. We check this phenomenon in different Monte Carlo samplings for different $d$ later.

For the set of 3d boxes in which we sample the base polytopes, we
generate a list of 4553 ``maximal'' boxes
$\mc{B}$ that can be thought as the dual of ``minimal''
$\mc{G}$-polytopes (without proving their completeness), defined as
\be
\mc{B}=\{v|\langle u,v\rangle\geq -6\ ,\ u\in\mc{G}\}\,.
\ee
The minimal $\mc{G}$-polytopes can be thought as
starting points of blow-ups in the minimal model of toric threefold
bases, which can be smooth\footnote{The list of these polytopes can be found on \href{https://doi.org/10.5281/zenodo.17761959}{https://doi.org/10.5281/zenodo.17761959} in 3d/3d-G-polytopes.txt.}. In this way, any base polytope $\Delta_b$
in the box $\mc{B}$ with origin contained in the interior of
$\Delta_b$ corresponds to a valid compact F-theory base at finite
distance in moduli space\footnote{For the cases with infinite
distance, see
\cite{Grimm:2018ohb,Lee:2021usk,Alvarez-Garcia:2023gdd,Alvarez-Garcia:2023qqj,Chen:2024cvc}
for recent developments using emergent extra dimensions or
strings.}.
A similar construction of ``maximal'' polytopes was used in
\cite{Avram:1997rs} to classify reflexive polytopes.

For detailed data analysis, we choose four different boxes denoted as $\mc{B}_6(p,q)$ that are (six times) dual to $\mc{G}$-polytopes with vertices $\{(1,0,0),(0,1,0),(0,0,1),(-1,-p,-q)\}$. One of them is the one containing the toric threefold base with the largest $h^{1,1}$\footnote{We compute $h^{1,1}$ of the toric base in this paper by taking the number of primitive lattice points $\vec{v}_i\in\Delta_b$, and then subtracting the number of linear relations, which is equal to the dimension of the base polytope.}, with vertices
\be
\mc{B}_6(84,516)=\{(3606,-6,-6)\ ,\ (-6,37,-6)\ ,\ (-6,-6,1)\ ,\ (-6,-6,-6)\}\,,
\ee
while the other three boxes are $\mc{B}_6(21,129)$, $\mc{B}_6(42,258)$ and $\mc{B}_6(63,387)$. The number of base polytopes inside each of these boxes is estimated to be $\sim 10^{85}$.

Given that the total number of 3d maximal boxes is around $\sim 5\times 10^3$, it is reasonable to give an upper bound estimation of the total number of 3d base polytopes as $\sim 10^{90}$. We compute the statistics of the Hodge
numbers and the number of rigid (automatically present) gauge groups
on threefold bases in these boxes,
including the ones with $E_7$ gauge groups that could potentially
realize an $E_7$ GUT model
following~\cite{Li:2021eyn,Li:2022aek,Li:2023dya}, as well as $E_8$
gauge groups that are the dominant rigid group containing the Standard
Model gauge group as a subgroup. Interestingly, the
total number of base polytopes in each of the four (rather different)
boxes
that we consider as example cases
are all of the same order of magnitude, despite  a large difference in
the sizes of these boxes. We also find that the non-Higgsable $E_7$ gauge
groups are more common in the Monte Carlo approaches in the boxes
$\mc{B}_6(21,129)$, $\mc{B}_6(42,258)$ and $\mc{B}_6(63,387)$, for
bases in the range of $h^{1,1}(B_3)\leq 140000$, while the 
number of $E_8$ factors is much greater in the larger boxes.

To 
confirm the validity of the statistical approach, we have also applied the Monte
Carlo program to the case of 2d base polytopes, in which we can
compare our results with a rigorous enumeration of such base
polytopes. In the 2d cases, the set of minimal $\mc{G}$-polytopes are
exactly $\mb{P}^2$, $\mb{F}_0$ and $\mb{F}_n$ $(12\geq n\geq 2)$,
which are the starting points of blow-ups for 2d bases of
F-theory~\cite{Grassi:1991tjo,Gross:1993fd,Morrison:2012js}. 
For 2d bases, the triangulation is unique, and
from this perspective, the set of allowed toric base polytopes is
larger than the set of 61539 2d bases in \cite{Morrison:2012js}. In
fact, this ensemble  includes smooth bases with certain type of
codimension-two (4,6) singularities in the elliptic fibration. We
estimate the total number of 2d base polytopes to be $N_{\rm
  tot,2d}\approx 2.9\times 10^9$. We also provide various consistency
checks in subsets, i.e. smaller 2d boxes defined relative to a
$k$-dual set of minimal polytopes, for which the complete set of 2d base polytopes can be enumerated.

The structure of the paper is as follows. In Section~\ref{sec:samp-box} we present the Monte Carlo setup in fixed boxes and the derivation of the redundancy factor used in all later computations. In Section~\ref{sec:canonical-box} we discuss how to generate the set of boxes corresponding to minimal $\mc{G}$ polytopes, in a canonical form. In Section~\ref{sec:Gaussian} we give a brief explanation of the Gaussian distribution of the number of vertices $m$ and the log of redundancy factor, observed in later data.

In Section~\ref{sec:2d} we apply the Monte Carlo approach to the case of $d=2$, and compare the results with certain well-defined subsets that can be exactly enumerated. We also analyze the statistics for the 2d base polytopes.

In Section~\ref{sec:3d} we present the most important $d=3$ case. We
first generate a (partial) 
list of minimal $\mc{G}$-polytopes that are
dual to the maximal 3d boxes used later, including
$\mc{G}$-polytopes with vertices
$\{(1,0,0),(0,1,0),(0,0,1),(-1,-p,-q)\}$ and the more general cases,
in Section~\ref{sec:3d-boxes}. In Section~\ref{sec:3d-largest} we
present and analyze the Monte Carlo data sampled in the largest box
$\mc{B}_6(84,516)$, and then do the same for three other boxes in
Section~\ref{sec:other-boxes}. We plot the average number of each kind
of non-Higgsable gauge group factors as a function of $h^{1,1}(B_3)$ in
Section~\ref{sec:gauge-group}, and then discuss the geometric and
physical meaning of the ensemble of 3d bases we get in
Section~\ref{sec:ensembles}.

Finally, we discuss the physical relevance of this analysis and future
directions in Section~\ref{sec:discussions}.

Various software programs and
  associated data files used to produce some of the results in this
  paper are included in a Zenodo archive \href{https://doi.org/10.5281/zenodo.17761959}{https://doi.org/10.5281/zenodo.17761959}.

\section{Random sampling of polytopes}
\label{sec:sampling}

In this section we go over the mathematical
details of the method that we use for
sampling polytopes.  While we are here sampling points on a lattice, a
similar approach has been used in the mathematical literature for sampling polytopes from a
continuous convex set; in particular, for such sampling certain
quantities satisfy a central limit theorem
\cite{vu2006central}
that helps illuminate the
distributions of polytopes that we encounter in the F-theory context.

\subsection{Sampling polytopes in a box}
\label{sec:samp-box}

The general combinatoric questions we would like to address are: ``What is the
total number of possible $d$-dimensional base polytopes associated
with compact toric $d$-folds that can support an elliptic fibration?,'' ``How can we randomly sample
them?,''   and ``What are their statistics?''. In this section we take the
dimension $d$ to be general.

The primary approach in this paper is to sample $n$ lattice points
$(v_1,v_2,\dots,v_n)$ in a $d$-dimensional box
$\mc{B}$, and check if the convex hull of the $n$ points
forms a valid base polytope
$\Delta_b$. We denote the total number of samplings by $N$. As the
vertices of a base polytope should be primitive rays, we only choose
each $v_i$ $(1\leq i\leq n)$ among the primitive rays in $\mc{B}$, and
denote by $|R|$ the number of possible choices for each $v_i$. There are thus
in total \be N_c=|R|^n \ee distinct possible choices for the set of
$(v_1,v_2,\dots,v_n)$. If there are $M$ valid base polytopes out of
the $N$ samplings, one would naively expect that the total number of
base polytopes should be $\sim N_c\cdot M/N$. However, there are a
variety of redundancy factors that should be divided out in the estimated
number of polytopes, due to the fact that many different choices of
$(v_1,v_2,\dots,v_n)$ would lead to the same polytope. In the
following discussion, we denote by $m$ the number of vertices for
$\Delta_b$.

The first and most significant redundancy factor, denoted by
$R(\Delta_b,m,n)$, describes the total number of ways to choose
$n$ points such that their convex hull forms the same polytope
$\Delta_b$ with $m$ vertices. When $m<n$, $(n-m)$ 
points 
out of $(v_1,v_2,\dots v_n)$
are  general primitive points
located inside the polytope $\Delta_b$, including the boundary
(and vertices) of
$\Delta_b$. In this case, there are additional redundancy factors due
to the fact that these $(n-m)$ 
points can be arbitrarily chosen inside
$\Delta_b$, and these different choices all give rise to the same
$\Delta_b$. We define $l(\Delta_b)$ to be the number of points inside
$\Delta_b$, including the boundary of $\Delta_b$.

To evaluate $R(\Delta_b,m,n)$, we first separate the $n$ points into $(m+j)$ points that are chosen among the $m$ vertices, and $(n-m-j)$ points that are chosen among the $(l(\Delta_b)-m)$ non-vertices. Here $j\in [0,n-m]$. The number of ways to choose the $(m+j)$ points among $n$ total points is $\binom{n}{m+j}$. Then after these $(m+j)$ points are chosen, we need to count the number of partitions of $(m+j)$ distinct objects into $m$ distinct sets, with at least one object in each set, we denote this number by $P(m+j,m)$. Finally, for the $(n-m-j)$ non-vertex points, there is a combinatoric factor of $(l(\Delta_b)-m)^{n-m-j}$. Finally summing over all possible values of $j$, we get the redundancy factor
\be
\label{red-factor}
R(\Delta_b,m,n)=\sum_{j=0}^{n-m}\binom{n}{m+j} P(m+j,m)(l(\Delta_b)-m)^{n-m-j}\,.
\ee
The quantity $P(m+j,m)$ can be computed as
\be
P(m+j,m)=\sum_{\text{Partitions}\ (a_1,\dots,a_m)\ \text{of}\ j}\frac{(m+j)!}{\prod_{i=1}^m(a_i+1)!}\,,
\ee
where $(a_1,\dots,a_m)$ $(a_i\geq 0)$ is an ordered partition of $j$. It is also equal to
\be
P(m+j,m)=S(m+j,m)\cdot m!\,,
\ee
where $S(m+j,m)$ is the Stirling number of second kind. The
combinatoric meaning of $S(m+j,m)$ is the number of different ways to
put $m+j$ different objects into $m$ identical boxes. An explicit
expression for $S(n,k)$ is 
\be
\label{stirling}
S(n,k)=\frac{1}{k!}\sum_{j=0}^{k}(-1)^{k-j}\binom{k}{j}j^{n}\,.
\ee

Now substituting (\ref{stirling}) into (\ref{red-factor}), we get
\be
R(l(\Delta_b),m,n)=\sum_{k=0}^{m}(-1)^{m-k}\binom{m}{k}k^m(l(\Delta_b-m)^{n-m})\times\sum_{j=0}^{n-m}\binom{n}{m+j}\left(\frac{k}{l(\Delta_b)-m}\right)^{j}\,.
\ee
Writing $p=j+m$, we have
\be
\label{any-v-R}
R(l(\Delta_b),m,n)=\sum_{k=0}^{m}(-1)^{m-k}\binom{m}{k}(l(\Delta_b)-m)^n\sum_{p=m}^{n}\binom{n}{p}\left(\frac{k}{l(\Delta_b-m)}\right)^p\,.
\ee

There is an approximate formula that simplifies (\ref{any-v-R}) when $l(\Delta_b)\gg n,m$.

We denote the $p$-th term on the r.h.s of (\ref{any-v-R}) $I(p)$, and consider the ratio of $I(p+1)$ and $I(p)$
\be
\frac{I(p+1)}{I(p)}=\frac{\binom{n}{p+1}}{\binom{n}{p}}\frac{k}{l(\Delta_b)-m}=\frac{k}{l(\Delta_b)-m}\frac{n-p}{p+1}\,,
\ee
which is a decreasing function of $p$, as well as $k$. The maximum of this ratio is reached at $k=p=m$, which equals to
\be
\left(\frac{I(p+1)}{I(p)}\right)_{\text{max}}=\frac{m}{l(\Delta_b)-m}\frac{n-m}{m+1}\,.
\ee
In the limit that $l(\Delta_b) \gg n >m $, an upper bound of the ratio can be given as 
\be
\frac{I(p+1)}{I(p)}<\frac{n}{l(\Delta_b)-m}\ll 1\,,
\ee
hence we find that the first term has the most contribution to the sum in (\ref{any-v-R}). As an example, take typical values in the 3d Monte Carlo approach in Section~\ref{sec:3d} $l(\Delta_b)=10^5,n=100,m=k=10$, a numerical calculation gives that
\be
\frac{\sum_{p=m}^{n}I(p)}{I(m)}=1.00082\,,
\ee
supporting the validity of the approximation in which we preserve the first term only.

In the above approximation, the redundancy factor can be written as
\be
R(l(\Delta_b),m,n)\approx (l(\Delta_b)-m)^{n-m}\binom{n}{m}\sum_{k=0}^{m}(-1)^{m-k}\binom{m}{k}k^m\,.
\ee
By another look, we realize that the sum over $k$ just gives an
expression of the factor $P(m,m)=m!\cdot S(m,m)$. Since there is only one
way to distribute $m$ objects into $m$ identical boxes, with at least
one object per box, we naturally
have $S(m,m)=1$. Finally, we arrive at a simple
approximate formula for the redundancy
factor \be
\label{Red-simple}
R(l(\Delta_b),m,n)=(l(\Delta_b)-m)^{n-m}m!\binom{n}{m}\,.
\ee

A rough heuristic explanation for
this formula can be understood as follows:
when $l(\Delta_b) \gg n$, the probability that one point is chosen
twice or more is suppressed by the factor $n/l(\Delta_b)$. Therefore,
in a good approximation, we can omit such contributions, and assume
that $n$ choices result in $n$ different points. In such 
circumstances, there are $m$ points chosen to be the vertices and
the remaining points lie in the other $l(\Delta_b)-m$ locations, and it is 
straightforward to check that $R=(l(\Delta_b)-m)^{n-m}m!\binom{n}{m}$.

The second independent redundancy factor comes from the possible $GL(d,\mb{Z})$ transformations $v_i'=M\cdot v_i$ $(i=1,\dots,n)$, such that the transformed vertices $v_i'$ $(i=1,\dots,n)$ still lie inside $\mc{B}$, and $v_i'$ is not a permutation of the old set of vertices $v_i$ (if it is a permutation, the redundancy is already taken into account in $R$). We denote the number of such $GL(d,\mb{Z})$ transformations, modding out the $S_n$ permutations, as $R_{GL(d,\mb{Z})}$. The estimated total number of $\Delta_b$ should be further divided by $1+R_{GL(d,\mb{Z})}$, due to the existence of $1+R_{GL(d,\mb{Z})}$ equivalent bases with different shapes in $\mc{B}$. After these two redundancy factors are taken into account, the estimated total number of bases in the Monte Carlo approach is
\be
\ba
\label{Ntot}
N_{\text{tot}}(n)&\approx \sum_{i=1}^M w_i\cr
&=\frac{1}{N}\cdot\left(\sum_{i=1}^M \frac{|R|^n}{R(l(\Delta_{b,i}),m_i,n)(1+R_{GL(d,\mb{Z})}(\Delta_{b,i}))}\right)\,,
\ea
\ee
where the sum is over valid base polytopes. We define the \textit{weight factor} associated to a base 
\be
\label{weight-1}
w_i\equiv\frac{|R|^n}{N\cdot R(l(\Delta_{b,i}),m_i,n)(1+R_{GL(d,\mb{Z})}(\Delta_{b,i}))}\,.
\ee

Furthermore, if there are multiple boxes $\mc{B}_1$, $\dots$, $\mc{B}_p$ to start from, and we take $N$ samplings for each box $\mc{B}_i$ $(1\leq i\leq p)$, when combining the data across the different boxes, it is possible that a base polytope $\Delta_b$ can fit into $t(\Delta_b)\geq 1$ boxes. In this case, the weight factor for each base should be divided by a third redundancy factor $t(\Delta_b)$:
\be
\label{weight-2}
w_i=\frac{|R|^n}{N\cdot R(l(\Delta_{b,i}),m_i,n)(1+R_{GL(d,\mb{Z})}(\Delta_{b,i}))t(\Delta_{b,i})}
\ee
and the total number of bases is then
\be
\ba
\label{Ntot-2}
N_{\text{tot}}(n)&\approx \sum_{i=1}^M w_i\cr
&=\frac{1}{N}\cdot\left(\sum_{i=1}^M \frac{|R|^n}{R(l(\Delta_{b,i}),m_i,n)(1+R_{GL(d,\mb{Z})}(\Delta_{b,i}))t(\Delta_{b,i})}\right)\,.
\ea
\ee

Finally, to get a more complete characterization of base polytopes, we often do $N$ samplings for $q\geq 1$ different values of $n$. In such cases, we combine the data for different $n$s, and simply replace the $N$ in the formula (\ref{Ntot}), (\ref{weight-1}), (\ref{weight-2}), (\ref{Ntot-2}) by $Nq$.

\subsection{Set of maximal boxes}
\label{sec:canonical-box}

An important question to address before setting up practical Monte Carlo samplings is how to choose appropriate boxes that can generate the complete set of $d$-dimensional base polytopes efficiently.

Imagine one starts from a very large box; a random polytope inside the
box would then likely fail to satisfy the $\mc{G}$-polytope condition, as
the $\mc{G}$ polytope is likely to be very small and not contain the
origin $\vec{0}$ in its interior. To increase  efficiency, in the
paper we use an alternative approach, such that we first try to
classify the different \textit{minimal $\mc{G}$ polytopes} for compact
$d$-dimensional toric bases. Then for each of these minimal $\mc{G}$
polytopes, we define a box $\mc{B}_6(\mc{G})$ that is defined as the six
times polar dual of $\mc{G}$; more generally, we define
\be
\mc{B}_k(\mc{G})=\{v\in\mb{Z}^d|\langle u,v\rangle\geq -k\ ,\forall
u\in\mc{G}\}\,.
\label{eq:bk}
\ee
The box $\mc{B}_6(\mc{G})$ then has the property that a polytope is contained
within the
box if and only if it satisfies the $\mc{G}$ condition for the associated
minimal $\mc{G}$ polytope, so the box is maximal for that
$\mc{G}$-polytope.

While for the purposes of determining bases for elliptic fibrations we
will principally focus on $\mc{B}(\mc{G})\equiv \mc{B}_6(\mc{G})$, in
various places we will also find it useful to consider the analogous
ensembles for other values of $k$.
For example, note that with this definition, a reflexive polytope is a lattice polytope $\nabla$ containing the origin as an interior point, with the property that $\mc{B}_1 (\mc{\nabla})$ is also a lattice polytope that contains the origin as an interior point.

The definition of a minimal $\mc{G}$ polytope consists of the following conditions:

\begin{enumerate}
\item $\vec{0}$ is in the interior of $\mc{B}(\mc{G})$, such that one can generate compact base threefolds from subsets of $\mc{B}(\mc{G})$.

\item $\vec{0}$ is in the interior of the dual polytope $\mc{G}$. When
  this is satisfied, for any base polytope $\Delta_b\subset
  \mc{B}(\mc{G})$, its $\mc{G}$-polytope
  $\mc{G}(\Delta_b)\supset\mc{G}$, which also contains $\vec{0}$ in
  the interior, and is automatically  at finite dinstance in F-theory moduli space.

\item After removing any single vertex of $\mc{G}$, and taking the
  remaining lattice points in $\mc{G}$, the new polytope does not
  satisfy the condition 2, hence the $\mc{G}$-polytope is really
  ``minimal''. This condition ensures that there is not another box
  $\mc{B}(\mc{G}')\supset \mc{B}(\mc{G})$, to avoid repetitions.

\end{enumerate}

In fact, the ``minimal condition'' in the condition 3 above exactly
coincides with the criteria for a minimal model in the toric minimal
model program, and $\mc{G}$ can be equivalently thought of as a toric
variety that cannot be blown down. Due to the condition 1, this toric
variety should also be a valid, possibly singular base of an elliptic
Calabi-Yau $(d +1)$-fold. 

In the 2d case, the complete set of $\mc{G}$s exactly corresponds to
the convex hulls of $\mb{P}^2$, $\mb{F}_0$, $\mb{F}_n$
$(n=2,\dots,12)$. These are precisely the minimal models for bases of
elliptic threefolds. For higher-dimensional cases, such a $\mc{G}$ may
have singularities if it is treated as a toric variety.
The set of toric minimal models for bases of elliptic fourfolds
is, as far as we know, as yet unknown;
the work here will, as a corollary, 
help to improve our understanding of this question.

Note that the minimal $\mc{G}$ polytopes defined similarly using
$\mc{B}_1$ contain as a subset the 16 minimal reflexive 3d polytopes
identified in \cite{Kreuzer:1998vb},  but also contain other non-reflexive minimal 3d polytopes such as the polytope with vertices $\{(1, 0, 0),
 (0, 1, 0), (0, 0, 1), (-1, -1, -2)\}$.

Finally, we comment on the canonical form of a $d$-dimensional
polytope or box with $m$ vertices, in which the $GL(d,\mb{Z})$
transformation redundancy has been removed. The key is that for any $d
\times m$ matrix $A$ with elements all integer, we can always find a
unique matrix $U \in GL(d,\mathbb{Z})$, such that $H=UA$ is a
row-style Hermite normal form matrix. A matrix $H$ 
has row-style Hermite normal form if the following properties hold:

\begin{enumerate}
\item $H$ is an upper triangular matrix, i. e. the matrix elements $H_{ij}=0$, for $i>j$.

\item Any rows of zeros are located below nonzero rows.

\item The first nonzero element of a nonzero row (also called pivot) is strictly to the right of the leading nonzero element of the row above it.

\item The elements above a pivot are non-negative, and strictly smaller than the pivot.

\end{enumerate}

For example, for a 3d polytope or box with $m\geq 4$ vertices, the canonical form of the vertices is
\be
\label{n-rays}
v_1=\bp 1 \\ 0 \\ 0\ep\ ,\ v_2=\bp p \\ q \\ 0\ep\ ,\ v_3=\bp r \\ s \\ t\ep\ ,\ v_k=\bp a_k \\ b_k \\ c_k\ep\quad (k=4,\dots,n)\,.
\ee
The integers satisfy $0\leq p< q$, $0\leq r<t$, $0\leq s<t$, while $a_k,b_k,c_k$ are unconstrained.

We are going to use this canonical form to describe the minimal $\mc{G}$-polytopes in the rest of the paper.

\subsection{Gaussian distributions}
\label{sec:Gaussian}

As will be seen in the detailed examples in the later sections, for
both cases of 2d and 3d, we  find that the $\log_{10}$ of the weight factor $w_i$ of $\Delta_{b,i}$ fits well to a Gaussian distribution. We provide a general qualitative argument as follows.

As derived in (\ref{weight-1}), (\ref{weight-2}), the weight factor of
a base is approximately equal to $\sim |R|^n/R(l(\Delta_b),m,n)\times
\mc{O}(10^{-1})$, where $|R|$ is the number of choices for each point,
$n$ is the total number of randomly chosen points, $l(\Delta_b)$ is
the number of primitive lattice points in the polytope $\Delta_b$,
and $m$ is the total number of vertices of $\Delta_b$. There are also
$(1+R_{GL(d,\mb{Z})})$ and $t(\Delta_{b,i})$ redundancy factors which
are often of the order $\mc{O}(10^{-1})$ in the later Monte Carlo
examples, which we will ignore in this discussion.

The dominant factor comes from $R(l(\Delta_b),m,n)$, which can be approximated by (\ref{Red-simple}) when $l(\Delta_b)\gg n$. We further take the limit of $n\gg m$, and estimate the $\ln$ of this factor by
\be
\ba
\ln(R(l(\Delta_b),m,n))&\approx(n-m)\ln (l(\Delta_b)-m)+n\ln n-n-(n-m)\ln(n-m)+(n-m)\cr
&\approx \text{const.}+n\ln(l(\Delta_b))-m\ln (l(\Delta_b))+m\ln n-m
 \label{eq:r-m}
\ea
\ee
after we absorb all the constant terms. One can also convert the formula into the $\log_{10}$ base.

One important variable, the number of vertices $m$, has a large effect
on $\ln(R(l(\Delta_b),m,n))$. As can be seen from examples, this
variable $m$ also approximately fits into a Gaussian
distribution\footnote{$n$ and $l(\Delta_b)$ should have effects as well, but the $n$ dependence should primarily affect the regime sampled, and not the net weights, and the range of variation in $ l(\Delta_b)/R \sim 1$ is not very wide either. Hence we primarily focus on the dependence on the single variable that is the number of vertices $m$.}. More rigorously, it has been proven in Theorem 1.2 of
\cite{vu2006central} that for an ensemble of random convex polytopes in a convex set
in $\mb{R}^d$, quantities such as the number of vertices obey a 
form of the central limit theorem and give a
 normal
distribution. We will see in Section~\ref{sec:2d} and
Section~\ref{sec:3d} two different Gaussian distributions 
that can be anticipated
following
this argument. First, the numbers of samplings in a Monte Carlo run obey a Gaussian 
distribution with respect to the number of vertices $m$. Second, the
numbers of samplings obey a Gaussian distribution with respect to the
$\log(w_i)$. 
 The first of these results can be seen as a discrete analogue  of the results in \cite{vu2006central}, while the second can be understood from
 (\ref{eq:r-m}), with the  variation largely dominated by  terms linear in
 $m$.

 Note that in our cases of random sampling of lattice polytopes, we
 are choosing among a discrete set of points instead of a continuous
 set of points. When the number of choices become small, there is
 expected to be deviation from the Gaussian distribution.   
 
As an amusing application, the Gaussian distribution of vertices for
polytopes can be also considered for the set of reflexive polytopes in the
Kreuzer-Skarke database~\cite{Kreuzer:2000qv,Kreuzer:2000xy}. The
473,800,776 reflexive polytopes naturally form a subset of the 4d
F-theory bases in the $k=1$ ensemble, subject to the additional
constraint that the dual of a reflexive polytope must also form a
lattice polytope. One
can check that the total number of 4d reflexive polytopes as a
function of the number of vertices also satisfy a Gaussian
distribution. It seems natural that the reflexive condition will
impose a stronger constraint on larger polytopes, but note that even an exponentially scaling constraint on the full set of polytopes would still give a net Gaussian
distribution, as observed.

\section{Monte Carlo approach for 2d base polytopes}
\label{sec:2d}

In order to confirm the validity of the main Monte Carlo approach used
in this paper, in this section we describe the analogous set of
polytopes for 2d bases (i.e.,  bases for elliptic Calabi-Yau
threefolds describing 6d F-theory models), where we can compare the
Monte Carlo results to exact calculations.
Note that, as in the case of base threefolds, these polytopes include
bases over which the elliptic fibration has certain types of (4, 6)
loci, and therefore goes beyond the list of toric 2d bases identified
in \cite{Morrison:2012js}.  Unlike in the case of base threefolds,
however, the triangulation of each 2d polytope is unique, so each of
these polytopes corresponds to a unique 2d toric variety.

\subsection{Set of all 2d base polytopes}
\label{sec:exact-2D}

 As mentioned before, the complete list of
minimal 2d $\mc{G}$-polytopes is
$\{\mb{P}_2,\mb{F}_0,\mb{F}_2,\dots,\mb{F}_{12}\}$. Hence we can compute
the corresponding dual boxes $\mc{B}_6(\mc{G})$ which are 6 times the
polar polytopes of $\mc{G}$. These boxes are listed in
Table~\ref{t:2Dboxes}.
We can generalize this set of boxes to other values of $k< 6$, as in
(\ref{eq:bk}).  For smaller values of $k$ the boxes only run up to
$\mc{B}_k(\mb{F}_{2k})$, and the vertices change correspondingly; for
example, for $k = 1$, the vertices of $\mc{B}_1 (\mb{F}_2)$ are $(-1,
1), (-1, -1), (3, -1)$.

\begin{table}
\centering
\begin{tabular}{|c|l|}
\hline
Box &  Vertices\\
\hline
$\mc{B}_6({\mb{P}^2})$ & $\{(-6,12),(-6,-6),(12,-6)\}$\\
$\mc{B}_6({\mb{F}_0})$ & $\{(6,6),(6,-6),(-6,6),(-6,-6)\}$\\
$\mc{B}_6({\mb{F}_2})$ & $\{(-6,6),(-6,-6),(18,-6)\}$\\
$\mc{B}_6({\mb{F}_3})$ & $\{(-6,4),(-6,-6),(24,-6)\}$\\
$\mc{B}_6({\mb{F}_4})$ & $\{(-6,3),(-6,-6),(30,-6)\}$\\
$\mc{B}_6({\mb{F}_5})$ & $\{(-6,2),(-4,2),(-6,-6),(36,-6)\}$\\
$\mc{B}_6({\mb{F}_6})$ & $\{(-6,2),(-6,-6),(42,-6)\}$\\
$\mc{B}_6({\mb{F}_7})$ & $\{(-6,1),(-1,1),(-6,-6),(48,-6)\}$\\
$\mc{B}_6({\mb{F}_8})$ & $\{(-6,1),(-2,1),(-6,-6),(54,-6)\}$\\
$\mc{B}_6({\mb{F}_9})$ & $\{(-6,1),(-3,1),(-6,-6),(60,-6)\}$\\
$\mc{B}_6({\mb{F}_{10}})$ & $\{(-6,1),(-4,1),(-6,-6),(66,-6)\}$\\
$\mc{B}_6({\mb{F}_{11}})$ & $\{(-6,1),(-5,1),(-6,-6),(72,-6)\}$\\
$\mc{B}_6({\mb{F}_{12}})$ & $\{(-6,1),(-6,-6),(78,-6)\}$\\
\hline
\end{tabular}
\caption{All the possible 2d boxes $\mc{B}_6(\mc{G})$ for 2d base polytopes with $k=6$, associated to the minimal $\mc{G}$-polytopes.}\label{t:2Dboxes}
\end{table}

For small values of $k$ we can exactly enumerate all polytopes lying
in the boxes $\mc{B}_k(\mc{G})$.
We have used two different approaches for such enumeration.

A
``top-down'' approach starts with the maximal polytope containing all $n$
primitive rays in a given box $\mc{B}$.  We then remove a single
vertex from this maximal polytope in all possible ways (and check that
the origin is included) to give all possible polytopes containing $n
-1$ primitive rays in $\mc{B}$.  We make a canonical ordering of this
list of polytopes, keeping each only once, and iterate to get the set
of allowed polytopes with $n -2$ primitive rays in $\mc{B}$.
We can then form the complete set of 2d polytope bases by taking the
union over the sets produced from the 13 distinct maximal boxes $\mc{B}$.

A ``bottom-up'' approach computes the set of all allowed polytopes that
fit in any of the boxes for a given $k$  in a single pass.  The idea is to start with all the 2d
minimal models ($\mb{P}^2$ and $\mb{F}_j, j \leq 2 k$), and blow up
points one at a time by adding new rays given by the sum of adjacent
rays, keeping all convex polytopes where the $k$-dual lattice polytope
contains the origin as an interior point.  Again, at each value of $n$
we keep a list of polytopes in canonical form.

\begin{table}
\centering
\begin{tabular}{|c|r|r|}
\hline
$k$ &  \# polytopes & max $h^{1,1} (B)$\\
\hline
1 & 16& 7\\
2 & 445 & 16\\
3 & 45430& 40\\
4 & 1306941 & 72\\
5 & 150493652& 140\\
\hline
\end{tabular}
\caption[x]{\footnotesize  The number of 2d polytope bases satisfying
  the $\mc{G}$-polytope condition for $k \leq 5$; for $k = 1$, the
  polytopes are the usual 16 reflexive 2d polytopes as listed in, e.g., \cite{Braun:2011ux,Klevers:2014bqa}.}
\label{t:2d-polytopes}
\end{table}

We have implemented both of these algorithms in Julia and run them on
a computing cluster for all values of $k \leq 6$\footnote{See the files on \href{https://doi.org/10.5281/zenodo.17761959}{https://doi.org/10.5281/zenodo.17761959} in the directory ``2d-code''.}.  For $k \leq 4$, both
algorithms run to completion, and we have checked that the results
agree. Since the top-down algorithm is run separately for each box,
while the bottom-up algorithm automatically includes all boxes, this
agreement is a strong check on the methodology.
For $k = 5$ the top-down algorithm also ran to completion;
the total number of polytopes for each value of $k \leq 5$ is given in
Table~\ref{t:2d-polytopes}.
For $k = 6$, the numbers of polytopes become quite large
($> 10^9$), and we have only run the computation from the bottom up to
$n \sim 30$ and from the top down to $n \sim 130$.  These results for
$k = 6$ and $k = 5$ are described in more detail in the following
subsections in comparison to the results of the Monte Carlo analysis.

\subsection{Monte Carlo sampling for 2d base polytopes}

In this section, we apply the sampling algorithm to probe the set of
all 2d base polytopes\footnote{The Mathematica code for the 2d Monte Carlo program is on \href{https://doi.org/10.5281/zenodo.17761959}{https://doi.org/10.5281/zenodo.17761959} in the file 2d/MonteCarlo-2d.nb.}.  For each 2d box, we take $N=100,000$ random
samplings for each $n\in\{10,20,30\}$. We find that for
$\mc{B}_6(\mb{P}^2)$, the $GL(2,\mb{Z})$ symmetry factor is typically equal to 5.  This follows because an
arbitrary polytope inside $\mc{B}_6(\mb{P}^2)$ can be transformed by
an element in $D_3$ while still remaining in the box
$\mc{B}_6(\mb{P}^2)$, while a typical polytope is big enough compared
to the box that there are no other transformations that keep it in the
box. The
$2\times 2$ matrix representation of these five $GL(2,\mb{Z})$
elements are \be \left\{\bp 0 & 1\\1 & 0\ep,\bp 0 & 1\\-1 & -1\ep,\bp
-1 & -1\\0 & 1\ep,\bp 1 & 0\\-1 & -1\ep,\bp -1 & -1\\1 &
0\ep\right\}\,.  \ee Similarly, $R_{GL(2,\mb{Z})}$  is generally equal
to 7 for $\mc{B}_6({\mb{F}_0})$, because of the $D_4$ symmetry of the
box $\mc{B}_6({\mb{F}_0})$. Moreover, $R_{GL(2,\mb{Z})}$ is generally
equal to 1 for $\mc{B}_6({\mb{F}_2})$ due to the $\mb{Z}_2$ symmetry
of $\mc{B}_6({\mb{F}_2})$, and usually equal to 0 for the other boxes
$\mc{B}_6({\mb{F}_n})$ $(12\geq n>2)$.

We list the estimated total number of 2d base polytopes in
Table~\ref{t:2D-box-results}, generated from each run with $N=100,000$
and a different $n$. As one can see, the total number of base
polytopes peaks around the box $\mc{B}_6({\mb{F}_9})$.  Note also that
the estimated number of polytopes is fairly insensitive to the choice
of $n$ in the range sampled\footnote{The Monte Carlo data are collected on \href{https://doi.org/10.5281/zenodo.17761959}{https://doi.org/10.5281/zenodo.17761959} in the file 2d/2d-k=6.zip.}.

\begin{table}[H]
\centering
\footnotesize
\begin{tabular}{|c|c|c|c|}
\hline
box & $n$ & $N$ & $N_{\text{tot}}$\\
\hline
\multirow{3}*{$\mc{B}_6({\mb{P}^2})$} & $10$ & $100000$ & $2.23\times 10^8$\\
 & $20$ & $100000$ & $2.36\times 10^8$\\
 & $30$ & $100000$ & $2.09\times 10^8$\\ 
 \hline
  \multirow{3}*{$\mc{B}_6({\mb{F}_0})$} & $10$ & $100000$ & $1.42\times 10^8$\\
 & $20$ & $100000$ & $1.28\times 10^8$\\
 & $30$ & $100000$ & $1.75\times 10^8$\\ 
 \hline
 \multirow{3}*{$\mc{B}_6({\mb{F}_2})$} & $10$ & $100000$ & $1.43\times 10^8$\\
 & $20$ & $100000$ & $1.99\times 10^8$\\
 & $30$ & $100000$ & $2.07\times 10^8$\\ 
 \hline
  \multirow{3}*{$\mc{B}_6({\mb{F}_3})$} & $10$ & $100000$ & $2.59\times 10^8$\\
 & $20$ & $100000$ & $3.51\times 10^8$\\
 & $30$ & $100000$ & $2.95\times 10^8$\\ 
 \hline
   \multirow{3}*{$\mc{B}_6({\mb{F}_4})$} & $10$ & $100000$ & $3.65\times 10^8$\\
 & $20$ & $100000$ & $4.09\times 10^8$\\
 & $30$ & $100000$ & $7.74\times 10^8$\\ 
 \hline
   \multirow{3}*{$\mc{B}_6({\mb{F}_5})$} & $10$ & $100000$ & $1.95\times 10^9$\\
 & $20$ & $100000$ & $7.23\times 10^8$\\
 & $30$ & $100000$ & $3.29\times 10^8$\\ 
 \hline
 \multirow{3}*{$\mc{B}_6({\mb{F}_6})$} & $10$ & $100000$ & $5.98\times 10^8$\\
 & $20$ & $100000$ & $1.06\times 10^9$\\
 & $30$ & $100000$ & $6.41\times 10^8$\\ 
 \hline
  \multirow{3}*{$\mc{B}_6({\mb{F}_7})$} & $10$ & $100000$ & $8.46\times 10^8$\\
 & $20$ & $100000$ & $1.36\times 10^9$\\
 & $30$ & $100000$ & $1.33\times 10^9$\\ 
 \hline
  \multirow{3}*{$\mc{B}_6({\mb{F}_8})$} & $10$ & $100000$ & $9.07\times 10^8$\\
 & $20$ & $100000$ & $1.37\times 10^9$\\
 & $30$ & $100000$ & $1.02\times 10^9$\\ 
 \hline
  \multirow{3}*{$\mc{B}_6({\mb{F}_9})$} & $10$ & $100000$ & $1.24\times 10^9$\\
 & $20$ & $100000$ & $2.15\times 10^9$\\
 & $30$ & $100000$ & $1.81\times 10^9$\\ 
 \hline
  \multirow{3}*{$\mc{B}_6({\mb{F}_{10}})$} & $10$ & $100000$ & $2.08\times 10^9$\\
 & $20$ & $100000$ & $1.72\times 10^9$\\
 & $30$ & $100000$ & $1.24\times 10^9$\\ 
 \hline
 \multirow{3}*{$\mc{B}_6(\mb{F}_{11})$} & $10$ & $100000$ & $7.54\times 10^8$\\
 & $20$ & $100000$ & $1.17\times 10^9$\\
 & $30$ & $100000$ & $1.22\times 10^9$\\ 
 \hline
 \multirow{3}*{$\mc{B}_6({\mb{F}_{12}})$} & $10$ & $100000$ & $1.05\times 10^9$\\
 & $20$ & $100000$ & $1.46\times 10^9$\\
 & $30$ & $100000$ & $1.03\times 10^9$\\ 
 \hline
\end{tabular}
\caption{2d Monte Carlo results for different boxes $\mb{B}_6({\mb{P}^2})$ and $\mb{B}_6({\mb{F}_n})$, $(n=0,2,3,\dots,12)$.}
\label{t:2D-box-results}
\end{table}

Now we combine the data from different boxes to form a comprehensive
scan of 2d base polytopes. Note that it is possible that a particular
base polytope $\Delta_{b,i}$ can fit into $t(\Delta_{b,i})\geq 1$
boxes, and we should use the formula (\ref{Ntot-2}) to estimate the
total number of bases. 
To compute the number $t(\Delta_{b,i})$, we take a box $\mc{B}_6(B)$ and multiply a matrix $M\in GL(2,\mb{Z})$ on each vertex $v\in \Delta_{b,i}$, and form a new lattice polytope $\Delta_{b,i}'=\{M\cdot v\}$. Then we explicitly check whether all the vertices of $\Delta_{b,i}'$ are contained in the box $\mc{B}_6(B)$. If so, it means that $\Delta_{b,i}$ can be contained in the box $\mc{B}_6(B)$ after the $GL(2,\mb{Z})$ transformation $M$. We add up the number of boxes $\mc{B}_6(B)$ satisfying this condition, and 
use this as
$t(\Delta_{b,i})$. Due to the fact that $GL(2,\mb{Z})$ is infinite, in practice we choose $M$ from the subset of $GL(2,\mb{Z})$ that are constructed as a product of no more than six generators of $GL(2,\mb{Z})$:
\be
\left\{\bp 1 & 0\\0 & -1\ep\ ,\ \bp 0 & 1\\1 & 0\ep\ ,\ \bp 1 & 1\\0 & 1\ep\right\},.
\ee
Hence it could potentially lead of an underestimation of $t(\Delta_{b,i})$ and an overestimation of the number of base polytopes.

Furthermore, after combining the data for
different $n$, the number of samplings $N$ in the denominator should
be replaced by $Nq=300,000$ in order to get correct numbers.

\begin{figure}
\centering
\subfloat{%
        \includegraphics[width=0.45\linewidth]{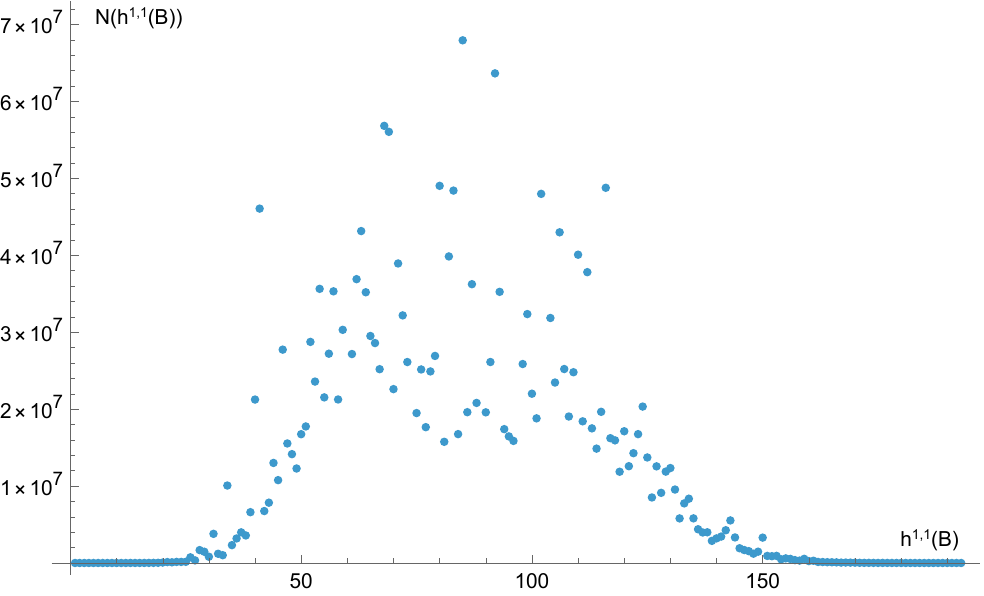}%
        }
        \hfill
        \subfloat{%
        \includegraphics[width=0.45\linewidth]{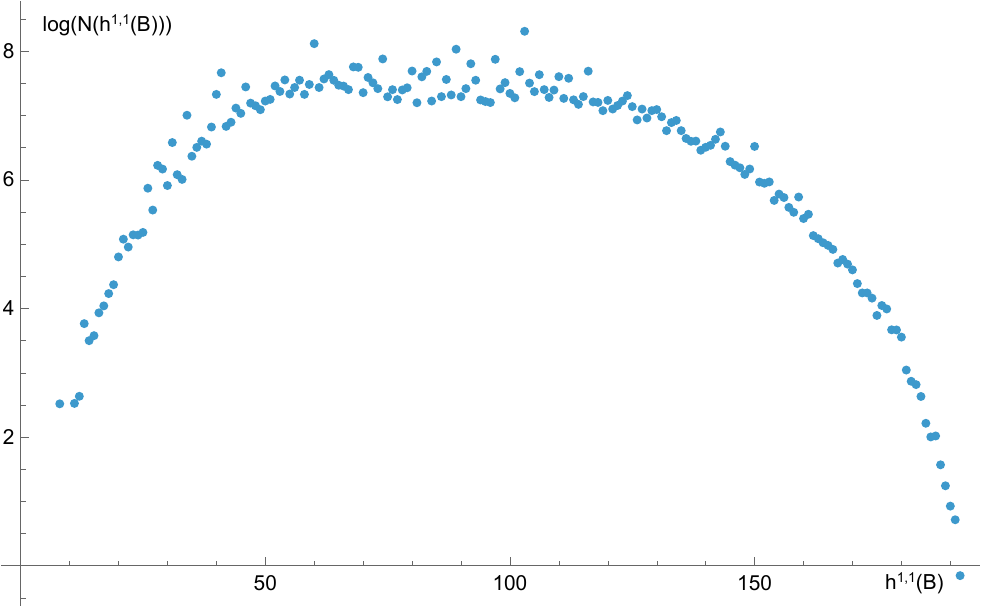}%
    }
    \\
    \subfloat{%
        \includegraphics[width=0.45\linewidth]{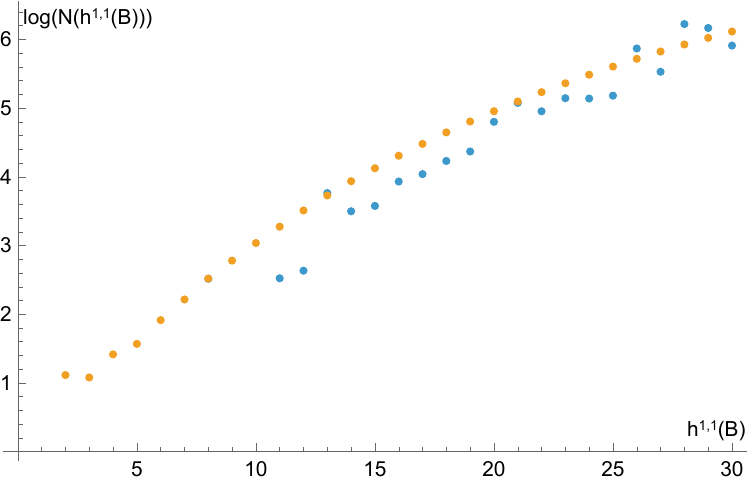}%
        }%
    \hfill%
        \subfloat{%
        \includegraphics[width=0.45\linewidth]{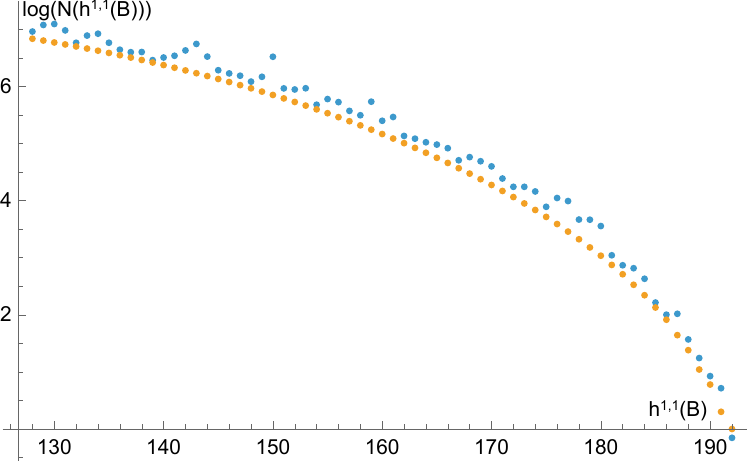}%
        }%
    \caption{The estimated number of 2d base polytopes for a given $h^{1,1}(B_2)$ and its $\log_{10}$, in blue color, with all the combined data across different boxes in Table~\ref{t:2D-box-results}. The regimes with small or large $h^{1,1}$ are magnified, for the $\log_{10}N_{\rm tot}$. The orange data points denote the exact number of 2d base polytopes for such $h^{1,1}(B_2)$ evaluated by the algorithm in Section~\ref{sec:exact-2D}.}\label{f:2dk6h11}
\end{figure}

After combining the data in Table~\ref{t:2D-box-results}, we get the estimated
total number of 2d base polytopes
\be
N_{\text{tot,2d}}\approx 2.9\times 10^9\,.
\ee
We plot the estimated number of base polytopes for a given
$h^{1,1}(B_2)$ in  Figure~\ref{f:2dk6h11}, with magnification
around small or large $h^{1,1}(B_2)$. In particular in these
subregions, the estimated numbers match well with the exact numbers of
bases with this $h^{1,1}(B_2)$, which strongly supports the
correctness of our Monte Carlo approach.\footnote{Note that the estimation is slightly too low for
small $h^{1,1} (B_2)$, and slightly large for large $h^{1,1} (B_2)$}

\begin{figure}
\centering
\includegraphics[height=7cm]{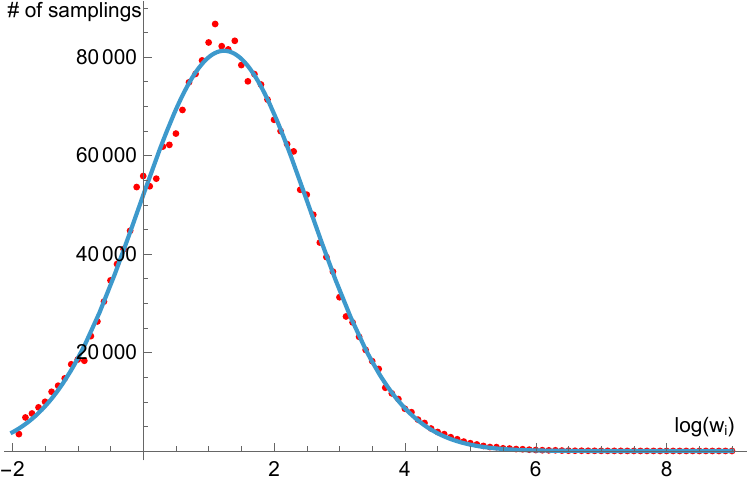}
\caption{The distribution of the weight factors $\log_{10}(w_i)$ in the Monte Carlo sampling of 2d base polytopes with $k=6$, across different boxes, with the data in Table~\ref{t:2D-box-results}. The horizontal axis is $\log_{10}(w_i)$, while the vertical axis is the number of samplings in each bin. The blue curve is the optimal fitting with a Gaussian function.}\label{f:2d-tot-weightfactor}
\end{figure}

\begin{figure}
\centering
\includegraphics[height=7cm]{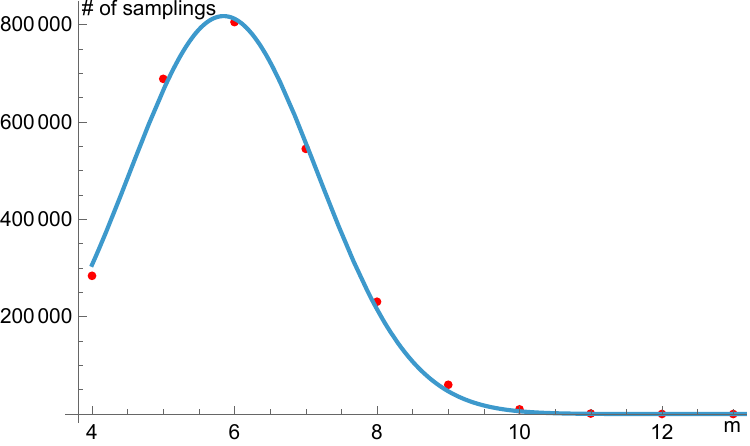}
\caption{The distribution of the number of vertices $m$ for each sample in the Monte Carlo sampling of 2d base polytopes with $k=6$, across different boxes, with the data in Table~\ref{t:2D-box-results}. The vertical axis denotes the number of samplings for each $m$. The blue curve is the optimal fitting with a Gaussian function.}\label{f:2d-vert-dist}
\end{figure}

 With this data,
we can confirm the Gaussian distribution of the $\log_{10}$ of weight
factors  discussed in Section~\ref{sec:Gaussian}. We plot in
Figure~\ref{f:2d-tot-weightfactor} the number of samplings for
different $\log_{10}(w_i)$, with bin size of $0.1$ on the horizontal
axis.
We also show the distribution of the number of vertices $m$ in Figure~\ref{f:2d-vert-dist}, which also approximately fits into a Gaussian distribution.

\begin{figure}
\centering
\subfloat{%
        \includegraphics[width=0.45\linewidth]{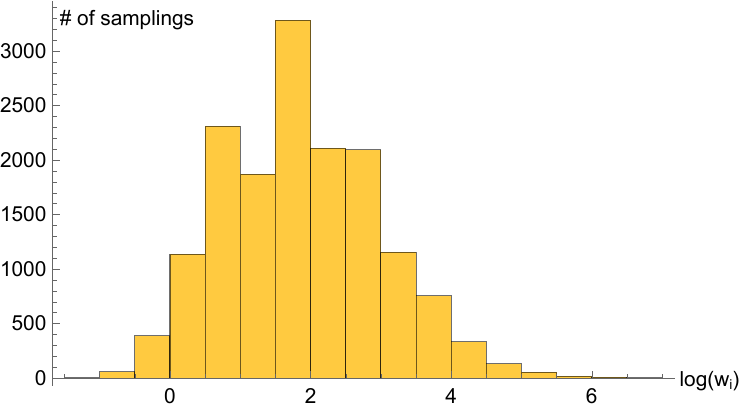}%
        }
        \hfill
        \subfloat{%
        \includegraphics[width=0.45\linewidth]{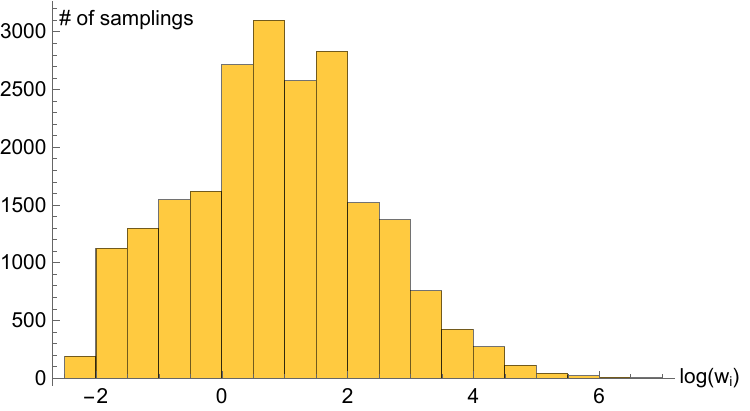}%
    }
    \\
    \subfloat{%
        \includegraphics[width=0.45\linewidth]{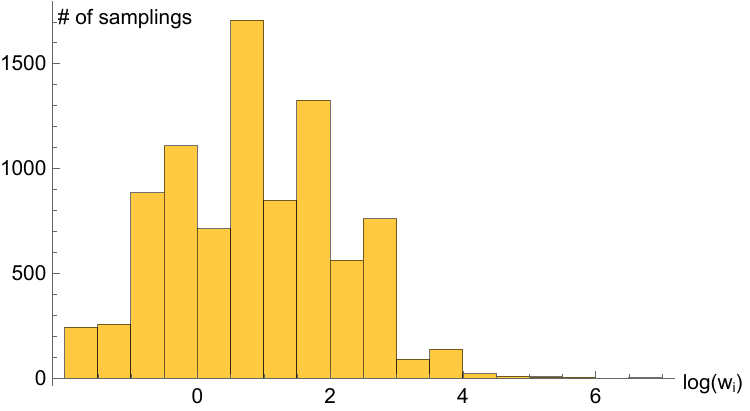}%
        }%
    \hfill%
        \subfloat{%
        \includegraphics[width=0.45\linewidth]{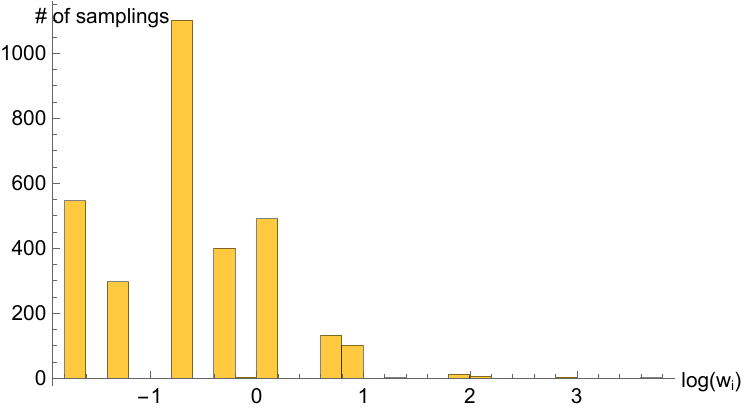}%
        }%
    \caption{The distribution of $\log_{10}(w_i)$ for the 2d Monte Carlo approach, for a fixed $h^{1,1}(B_2)$. The top-left, top-right, bottom-left and bottom-right figures correspond to $h^{1,1}(B_2)=50$, $100$, $150$ and $180$ respectively.}\label{f:2d-weightfactors}
\end{figure}

Furthermore, we plot the distribution of the log of the weight factors
$\log_{10}(w_i)$ among polytopes with a fixed
$h^{1,1}(B_2)\in\{50,100,150,180\}$, in
Figure~\ref{f:2d-weightfactors}. One can see a large deviation from
 normal  (Gaussian) form for larger $h^{1,1}(B_2)$, due to sample
 limitation,  i.e., the fact that the
number of distinct base polytopes in that range is small.

 \subsection{Consistency checks}
 
We present further consistency checks for the 2d Monte Carlo approach. First, let us sample base polytopes in a very small box of $\mc{B}=\{(-1,2),(2,-1),(-1,-1)\}$, which contains 14 of the 16 2d reflexive polytopes. This box $\mc{B}$ contains 9 primitive rays.

Thus the estimated total number of bases is computed by
\be
\ba
N_{\text{tot}}(n)\approx &\frac{1}{N}\cdot\sum_{i=1}^M w_i\cr
=&\frac{1}{N}\cdot\left(\sum_{i=1}^M \frac{9^n}{R(l(\Delta_{b,i}),m,n)(1+R_{GL(2,\mb{Z})})}\right)\,.
\ea
\ee

We take the sampling size of $N=10000$ and the number of points $n=10$, and we get an estimation $N_{\text{tot}}(10)\approx 14.3$, which is close to the actual number of reflexive polytopes---14 in the box $\mc{B}$.

We also try to generate the complete subset of 2d base polytopes in a ``$k=5$'' set of smaller 2d boxes, which are given by the Newton polytope of $-5K_B$ for $B=\mb{P}_2,\mb{F}_0,\mb{F}_n\ (n=2,\dots,10)$, listed in table~\ref{t:2Dboxes-k5}, denoted as $\mc{B}_{5}(B)$.

\begin{table}
\centering
\begin{tabular}{|c|l|}
\hline
Box &  Vertices\\
\hline
$\mc{B}_{5}({\mb{P}^2})$ & $\{(-5,10),(-5,-5),(10,-5)\}$\\
$\mc{B}_{5}({\mb{F}_0})$ & $\{(5,5),(5,-5),(-5,5),(-5,-5)\}$\\
$\mc{B}_{5}({\mb{F}_2})$ & $\{(-5,5),(-5,-5),(15,-5)\}$\\
$\mc{B}_{5}({\mb{F}_3})$ & $\{(-5,3),(-4,3),(-5,-5),(20,-5)\}$\\
$\mc{B}_{5}({\mb{F}_4})$ & $\{(-5,2),(-3,2),(-5,-5),(25,-5)\}$\\
$\mc{B}_{5}({\mb{F}_5})$ & $\{(-5,2),(-5,-5),(30,-5)\}$\\
$\mc{B}_{5}({\mb{F}_6})$ & $\{(-5,1),(-1,1),(-5,-5),(35,-5)\}$\\
$\mc{B}_{5}({\mb{F}_7})$ & $\{(-5,1),(-2,1),(-5,-5),(40,-5)\}$\\
$\mc{B}_{5}({\mb{F}_8})$ & $\{(-5,1),(-3,1),(-5,-5),(45,-5)\}$\\
$\mc{B}_{5}({\mb{F}_9})$ & $\{(-5,1),(-4,1),(-5,-5),(50,-5)\}$\\
$\mc{B}_{5}({\mb{F}_{10}})$ & $\{(-5,1),(-5,-5),(55,-5)\}$\\
\hline
\end{tabular}
\caption{The restricted ``$k=5$'' set of 2d boxes for 2d base polytopes generating a subset of base polytopes that can be completely enumerated.}\label{t:2Dboxes-k5}
\end{table}

\begin{figure}
\centering
\subfloat{%
        \includegraphics[width=0.45\linewidth]{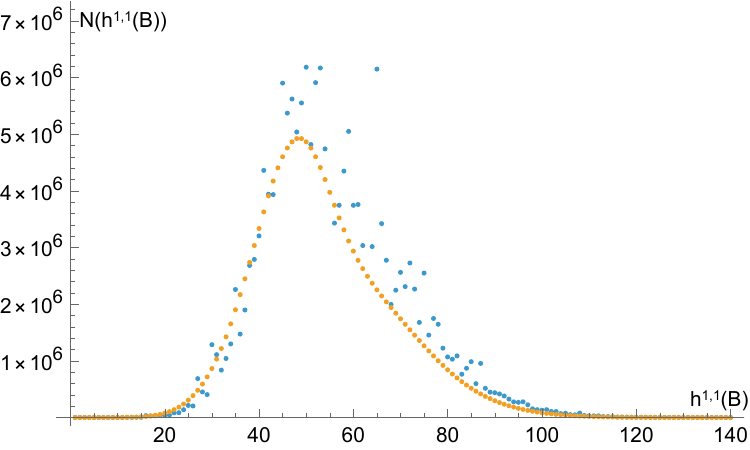}%
}
\hfill
\subfloat{%
        \includegraphics[width=0.45\linewidth]{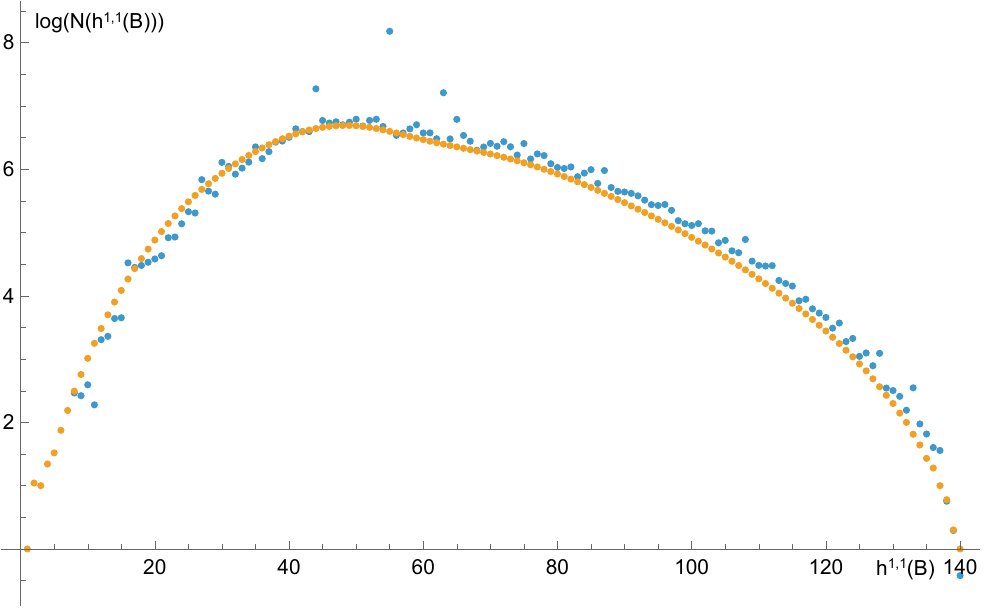}%
    }
    \\
    \subfloat{%
        \includegraphics[width=0.45\linewidth]{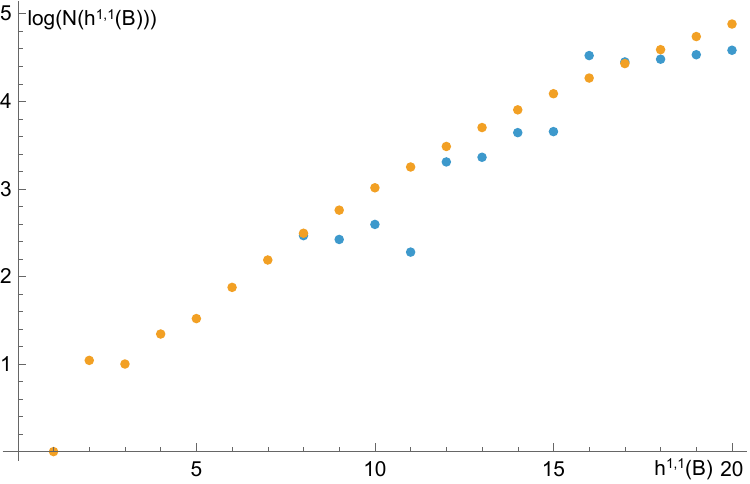}%
        }%
    \hfill%
        \subfloat{%
        \includegraphics[width=0.45\linewidth]{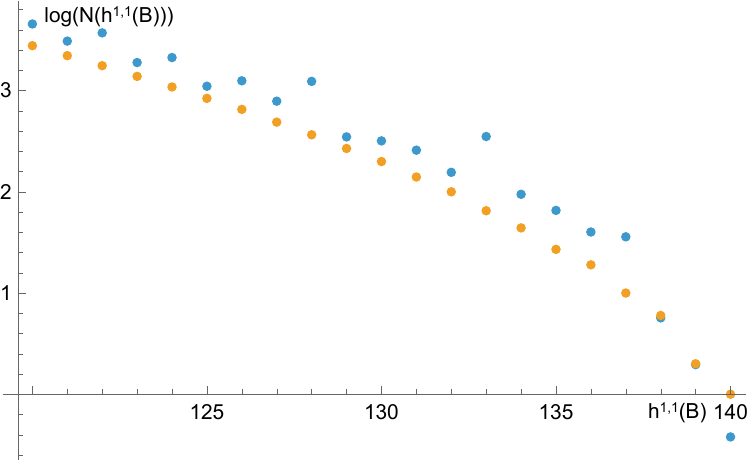}%
        }%
    \caption{The estimated number of 2d base polytopes for a given
      $h^{1,1}(B_2)$ and its $\log_{10}$,  for the $k = 5$
case,
with all the combined data across different boxes in Table~\ref{t:2Dboxes-k5}. We have matched the numbers from the Monte Carlo program (in blue) with the exact numbers (in orange). The regimes with small or large $h^{1,1}(B_2)$ are magnified.}\label{f:2dk5h11}
\end{figure}

Using the same methods as in the last section to combine the different data from different boxes (\ref{Ntot-2}), we estimate the total number of base polytopes fitting in these boxes to be\footnote{The Monte Carlo data are collected on \href{https://doi.org/10.5281/zenodo.17761959}{https://doi.org/10.5281/zenodo.17761959} in the file 2d/2d-k=5.zip.}
\be
N_{\rm tot}\approx 2.5\times 10^8\,.
\ee
 This matches (within a factor of 2) the exact number of $\approx
 1.5\times 10^8$ from Table~\ref{t:2d-polytopes}.
In this subset, we also plot the estimated vs.\  exact number of base polytopes
for a given $h^{1,1}(B_2)$ in Figure~\ref{f:2dk5h11}, with
magnification around small and large $h^{1,1}$, with good agreement.

\begin{figure}
\begin{center}
\includegraphics[width=12cm]{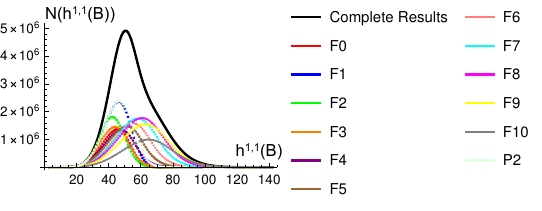}
\end{center}
\caption[x]{\footnotesize The distribution of allowed polytopes for
  the different $k = 5$ boxes; note that some polytopes are contained
  within multiple boxes so the total curve is less than the sum of the
parts.}
\label{f:all-legend}
\end{figure}


In Figure~\ref{f:all-legend}, we compare the exact number of base polytopes for each $h^{1,1}$ with the number inside each $k=5$ boxes. The combined curve is not a simple addition of the curves from different boxes, because as explained before, a base can generally fit into more than one box. 
It is interesting to note that the distribution of polytopes is not
dominated by one or two of the boxes, even though some boxes are much
larger than others, in the sense of containing more primitive lattice points.

\begin{figure}
\centering
\subfloat{%
        \includegraphics[width=0.45\linewidth]{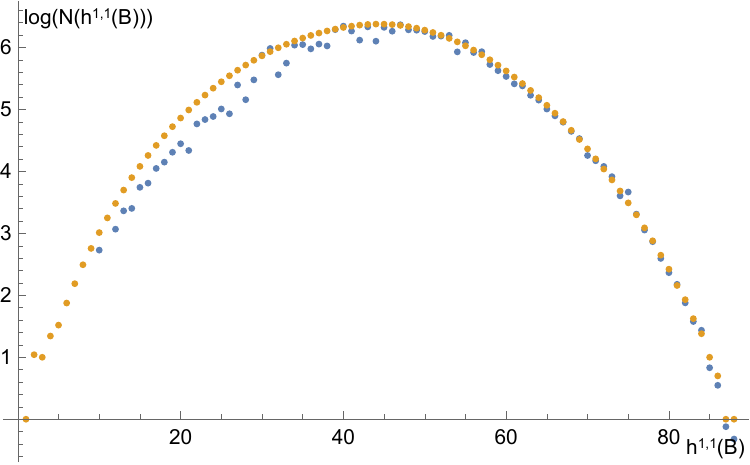}%
}
\hfill
\subfloat{%
        \includegraphics[width=0.45\linewidth]{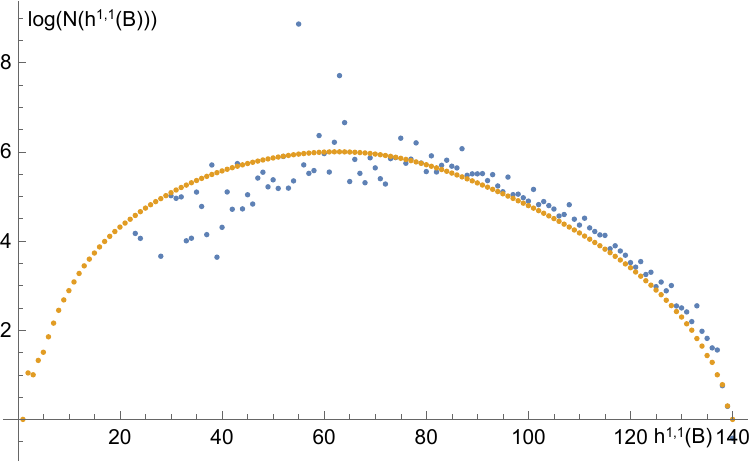}%
    }
   
    \caption{The $\log_{10}$ of the estimated number of 2d base polytopes for a given
      $h^{1,1}(B)$, for the $k = 5$
case, for the boxes $\mc{B}_5({\mb{P}^2})$ (on the left) and $\mc{B}_5({\mb{F}_{10}})$ (on the right). We have matched the numbers from the Monte Carlo program (in blue) with the exact numbers (in orange).}\label{f:k=5-diff-box}
\end{figure}

In Figure~\ref{f:k=5-diff-box}, we also show the ($\log_{10}$ of) number of base polytopes inside a particular box, e.g. $\mc{B}_{5}(\mb{P}^2)$ and $\mc{B}_{5}(\mb{F}_{10})$. The Monte Carlo result again matches the exact numbers with a very high accuracy.

\subsection{Different base ensembles}
\label{sec:2d-ensembles}

From  this analysis, we have identified a larger set of 2d base
polytopes than the 61539 toric bases~\cite{Morrison:2012js}. In this
section, we carefully compare  three distinct possible ensembles of 2d bases:

\begin{itemize}
\item (I) The set of 61539 toric bases, which are smooth and do not
  have codimension-two (4,6) loci at the intersection points of toric
  curves~\cite{Morrison:2012js}. Such bases are allowed to have
  $(-9)$, $(-10)$ or $(-11)$-curves with non-toric codimension-two
  (4,6) loci, corresponding to rank-1 E-string subsectors. One can
  blow up these codimension-(4,6) points to get a  (non-toric) base with only
  $(-12)$-curves and $(-n)$-curves ($n\leq 8$) as the Mori cone
  generators.

\item (II) The set of 2d base polytopes $\Delta_b$ in our ensemble
  analyzed in this paper. After we add all the primitive internal
  points in $\Delta_b$ as toric rays of the toric base $B_2$, $B_2$ is
  always smooth. Thus each $\Delta_b$ corresponds to a unique smooth
  2d base constructed in this way.

Nonetheless, these smooth bases would generally have cod-2 (4,6) loci at the intersection points of toric curves, for example the local
configuration  shown in Figure~\ref{f:2D-new}. This is a $(-4)-(-1)-(-5)$ curve
configuration, which contains codimension-two $(4,6)$ loci that encode
6d (1,0) SCFT sectors. After blowing up these $(4,6)$ points, the
newly added rays would be located outside of the line $(3,1)-(-1,-4)$, and
the convex hull would be changed.

In fact, these 2d bases in ensemble (II) can be  generated from
blowing down $(-1)$-curves from the 61539 bases in the ensemble
(I). The bases are kept smooth during this process. From the Monte
Carlo program, we estimate that there are  
$ \sim 3\times 10^{9}$
bases in the
ensemble (II). 

However,  each base polytope $\Delta_b$ in the ensemble (II)
can contain more than one smooth 2d base whose convex hull is
$\Delta_b$. For example in the local curve configuration of
$(-4)-(-1)-(-5)$, the $(-1)$-curve in the middle is in the interior of
$\Delta_b$, and can be contracted to get a local curve configuration
of $(-3)-(-4)$, which is a different smooth base.

\begin{figure}
\centering
		\begin{tikzpicture}
			\node (1) at (0,0) {};
                \node (2) at (3,1) {};
                \node (3) at (1,0) {};
                \node (4) at (0,-1) {};
                \node (5) at (-1,-4) {};
                \node (6) at (1,-1) {};
                \node (a2) at (3.5,1) {$(3,1)$};
                \node (a3) at (1.5,0) {$(1,0)$};
                \node (a4) at (0.2,-1.2) {$(0,-1)$};
                \node (a5) at (-2,-4) {$(-1,-4)$};
                \node (a6) at (1.3,-1.2) {$(1,-1)$};
                \draw[->] (1) to (2);
                \draw[->] (1) to (3);
                \draw[->] (1) to (4);
                \draw[->] (1) to (5);
                \draw[->] (1) to (6);
                \draw (2) to (5);
            \end{tikzpicture}
            \caption{A local picture of an allowed 2d base
              polytope. After adding all the primitive rays in the
              interior, the base contains codimension-two $(4,6)$
              singularities in the elliptic fibration over toric
              points,
 and hence does not fit into the classification of \cite{Morrison:2012js}.}\label{f:2D-new}
\end{figure}
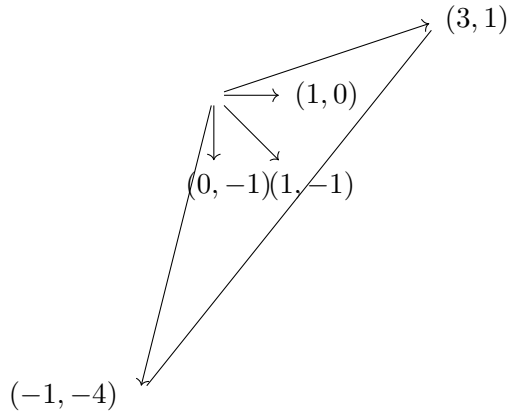

\item (III) The full set of 2d bases with cod-2 (4,6) loci, which are
  not necessarily smooth. All these bases can be obtained by removing
  an arbitrary number of primitive internal points in a 2d base
  polytope $\Delta_b$. For each $\Delta_b$ with number of primitive
  internal points $l(\Delta_b)$, one can construct $2^{l(\Delta_b)-m}$ 
  bases from it, by choosing whether to keep or remove each primitive point in the interior of $\Delta_b$. From the 2d Monte Carlo result, we estimated the
  total number of bases in the ensemble (III) to be $\sim 10^{57}$
  (the largest box has 191 internal points).

\end{itemize}

For the class (II), we analyze the $(4,6)$ points  in the 2d bases
that arise
from adding all the primitive internal points in the base polytopes
$\Delta_b$ generated in our program. In order to classify the
corresponding 6d (1,0) SCFTs  \cite{HeckmanMorrisonVafa}, we compare
the rays $\nu_i$ on the toric 
base $B_2$ with $(4,6)$ points with the rays $\nu_i'$ on the toric
base $B_2'$ where all $(4,6)$ points are blown up. The toric polytope
of the base $B_2'$ can be effectively generated by taking the double
dual of $B_2$, i.e. first define \be
\mc{G}(B_2)=\{u\in\mb{Z}^2|\langle u,\nu_i\rangle\geq
-6\ ,\ \forall\nu_i\in B_2\}\,, \ee and then generate the toric
polytope of $B_2'$ as \be \Delta_b'=\{\nu'\in\mb{Z}^2|\langle
u_i,\nu'\rangle\geq -6\ ,\ \forall u_i\in \mc{G}(B_2)\}\,.  \ee We
 can list all the rays $\nu_i$ of $B_2$ and $\nu_i'$ of $B_2'$ in 
counter-clockwise order. Then one tries to find the rays on $B_2'$
that correspond to the neighboring rays on $B_2$, i.e. $\nu_j'=\nu_i$
and $\nu_k'=\nu_{i+1}$. If $k>j+1$, which means that there are rays
between $\nu_j'$ and $\nu_k'$, such rays would give rise to the tensor
branch of the 6d (1,0) SCFT located between $\nu_i$ and $\nu_{i+1}$ on
the original base $B_2$.

Using this algorithm, we find that the 6d (1,0) SCFTs
that arise in our Monte Carlo ensemble all fit into the following three classes:
\begin{enumerate}
\item Rank-1 E-string with tensor branch $1$ (i.e., a single
  additional -1 curve).
 
\item Rank-2 E-string with tensor branch $2-1$.

\item A rank-3 SCFT with tensor branch
\be
\overset{\mathfrak{su}(2)}{2}-2-1\,.
\ee
\end{enumerate}

In conclusion, our ensemble (II) only counts the number of compact
smooth toric bases with the aforementioned three kinds of 6d (1,0)
SCFTs. There are certainly other smooth bases with more general 6d
(1,0) SCFTs, that can be generated from blowing down $(-1)$-curves
from one of our base polytopes $\Delta_b$. 
 It is not clear to us whether there is physical significance to this
 particular subset of SCFTs;  this is an interesting question for
 further work.

\section{Monte Carlo approach for 3d base polytopes}
\label{sec:3d}

Having described the algorithm in some detail and  having  verified the
accuracy against exactly computed data for 2d polytopes, we now
analyze the set of 3d base polytopes.  We begin by analyzing the set
of minimal $\mc{G}$-polytopes that give rise to canonical maximal
boxes $\mc{B}$.  We then carry out Monte Carlo calculations in the
largest box and several other boxes, collect data on gauge groups and
other features, and summarize the different base ensembles that may be
relevant for toric 3d bases.

\subsection{Set of minimal $\mc{G}$-polytopes}
\label{sec:3d-boxes}

Following  the logic of Section~\ref{sec:canonical-box}, we investigate the
conditions on minimal 3d $\mc{G}$-polytopes. For a polytope to be minimal,
after removing any 
vertex, the remaining polytope has to be non-compact,
i.e. the origin $(0,0,0)$ would not be in the interior any more. For this
condition to be satisfied,  a simple geometric argument indicates that the number of
vertices in a minimal $\mc{G}$-polytope cannot exceed 6, and the only
minimal $\mc{G}$-polytope with 6 vertices is 
$\mb{P}^1\times\mb{P}^1\times\mb{P}^1$ with vertices \be
\{(1,0,0),(0,1,0),(0,0,1),(-1,0,0),(0,-1,0),(0,0,-1)\}\,, \ee as this
is the only possible polytope with 6 vertices that becomes
non-compact after one removes any of its vertices. 
\footnote{To see the upper bound of 6, we triangulate the boundary of
the polytope and consider the solid angle subtended by
the faces surrounding each vertex.  A vertex can be removed while
maintaining the compactness property as long as this solid angle is
less than $2 \pi$.  Thus, for a  polytope with 7 or more vertices, none
of which can be removed, summing the solid angles of the adjacent
faces over all vertices gives $>14 \pi$. Each face is included 3
times, so the total solid angle would be $> 14/3 \pi > 4 \pi$, a
contradiction.  For 6 vertices, the bound of $4 \pi$ is exactly
saturated only when the sum of the adjacent spaces is exactly $2 \pi$,
which occurs only for the stated 6-vertex (octahedral) polytope.}

We  thus wish to enumerate the $\mc{G}$ polytopes with 4 and 5 vertices. As in Section~\ref{sec:canonical-box}, we transform each of these $\mc{G}$ polytopes into the Hermite normal form.

A particularly important subset is the
set of  $\mc{G}$ polytopes  with the following 4 vertices:
\be
\{(1,0,0)\ ,\ (0,1,0)\ ,\ (0,0,1) ,\ (-1,-p,-q)\}\,;
 \label{eq:special-g}
\ee
 we denote the corresponding box by $\mc{B}_6(p,q)$. 

\begin{figure}
\centering
\includegraphics[height=5cm]{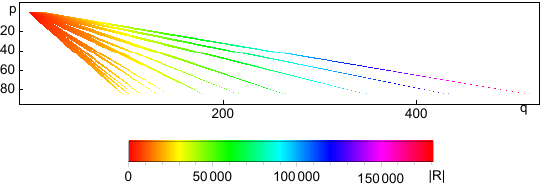}
\caption{Distribution of $p, q$ values for minimal $\mc{G}$ polytopes
of the form (\ref{eq:special-g}),
  and the number of primitive points $|R|$ for the corresponding base polytopes. The horizontal
  and vertical axis stands for values of $q$ and $p$
  respectively. Each point denotes a 3d minimal $\mc{G}$-polytope,
  while the color labels the number of primitive rays in the
  corresponding $\mc{B}_6(p,q)$.}\label{f:heat} 
\end{figure}

There are strict bounds on the parameters $p$ and $q$: $p \leq 84$ and
$q \leq 516$, in order for the $\mc{B}_6(p,q)$ to be
compact. Moreover, we require $p \leq q$ to avoid  the redundancy of
switching $p$ and $q$. Since we have vertices with $x=1$ and $x=-1$,
and there are no edges perpendicular to the $yz$ plane, we conclude that
the integer points inside the polytope are all located at $x=0$ except
two vertices. However, we cannot form a good polytope containing the
origin in the interior without points with negative $x$ coordinate
after removing the vertex $(-1,-p,-q)$. Therefore we see that the
conditions on $\mc{G}$ are equivalent to $\mc{G}$ being a valid base
polytope. By explicit enumeration, we find all minimal $\mc{G}$-polytopes
of this type, as well as the boxes $\mc{B}_6(p,q)$. We plot the values
of $(p,q)$ for these solutions, as well as the number of primitive
rays in the corresponding $\mc{B}_6(p,q)$, in Figure~\ref{f:heat}.

In particular, the $\mc{G}$ polytope with the largest values
$(p,q)=(84,516)$ exactly corresponds to the $\mc{G}$-polytope of the
base threefold $B_3$ with the largest $h^{1,1}(B_3)$, studied in
\cite{Wang:2020gmi}. The corresponding box $\mc{B}_6(84,516)$ has
vertices
\be
\label{maxver}
\mc{B}_6(84,516)=\{(3606,-6,-6)\ ,\ (-6,37,-6)\ ,\ (-6,-6,1)\ ,\ (-6,-6,-6)\}\,,
\ee
and has $|R|=181203$ primitive lattice points. This largest box will be the main subject of research in the next section.

A natural question is: can we
completely determine those 
maximal boxes that are not
of 
the above $\mc{B}_6(p, q)$ type. 
From the
experience above with the  $\mc{B}_6(p, q)$ distribution, it seems plausible that as
long as we fix the first three vertices, the distribution of the
fourth vertex  will be  fairly contiguous, or at most  will have
 relatively small
jumps  between valid $\mc{G}$-polytopes. Keeping this idea in mind, we
propose the following algorithm to search
for the set of 3d minimal $\mc{G}$ polytopes (or at least the outline
of the minimal box set). Here we take the case of 4 vertices as an
example:

\begin{enumerate}
\item Choose a bound $A$ for the first three vertices, then generate the full set of first three vertices in the bounded region. In other words, we first find all possible triples 
\be
\{(1, 0, 0), (p, q, 0), (r, s, t)\}
\ee
in the Hermite normal form (\ref{n-rays}) with $0 < q$, $t \leq A$, $0 \leq p <q$, $0 \leq r,s < t$. Since we are considering $\mc{G}$ polytopes here, we do not require the vertices of the $\mc{G}$ polytope to be primitive. However, the $\mc{G}$ polytopes with non-primitive vertices are not minimal in general as one can imagine.

\item Fix a choice of first three vertices, then choose a big enough
  bound $A_4$, and enumerate all possible fourth vertices $(-a, -b, -c)$
  with $0 < a, b, c \leq A_4$. Pick all the minimal cases as the
  initial set for the next step.

\item  Then we try to identify further minimal $\mc{G}$ polytopes near the minimal set we have found. That is, we keep the first three vertices unchanged, and replace the fourth vertex by points near it. For example, if we know the $\mc{G}$ polytope\be
\{(1,0,0),(p,q,0),(r,s,t),(-a,-b,-c)\}
\ee
is minimal, we will examine and pick out the minimal $\mc{G}$ polytopes from the set of all the $\mc{G}$ polytopes with the form
\be
\{(1,0,0),(p,q,0),(r,s,t),(-a-\Delta a,-b-\Delta b,-c-\Delta c)\},
\ee
where $0\leq\Delta a , \Delta b , \Delta c <\Delta$, and $\Delta$ is a step size we choose in advance. Here we keep $\Delta a, \Delta b, \Delta c$ nonnegative from the knowledge we get in Figure~\ref{f:heat} that new minimal $\mc{G}$ polytopes would appear in the negative direction from smaller minimal $\mc{G}$ polytopes. We collect all the minimal $\mc{G}$ polytopes we find as the new initial minimal set, and repeat the progress before until we cannot find new minimal $\mc{G}$ polytopes. Then we assume that we have come to an end of the set of minimal $\mc{G}$ polytopes (for a chosen
  combination of the first three vertices chosen in the first step.) 
\end{enumerate}

By expanding the bound $A$, we can detect the change of our minimal
set. If we find no change after a big enough expansion, we can
conclude that the set of minimal boxes have likely been identified.

Note that the set of minimal $\mc{G}$ polytopes found in this way
still has the permutational redundancy described in previous sections. In
our enumeration, we first ignore the possible permutational
redundancy, then keep only $\mc{G}$ polytopes that cannot transform to
each other after permutation of vertices and $GL(3, \mb{Z})$
transformation.

We apply our algorithm on polytopes with $A=12$, $A_4=20$, $\Delta = 6$. In other
words, we enumerate $\mc{G}$ polytopes with the first three vertices
containing only elements less than or equal to 12 and the fourth vertex with
elements that are all negative but greater than $-20$, and explore around the
minimal $\mc{G}$ polytopes already found. After modding out
permutational redundancy,  a total of 314 different $\mc{G}$ polytopes other than
the $\mc{B}_6(p, q)$ type are found to be minimal. The explicit list of minimal $\mc{G}$ polytopes can be found on \href{https://doi.org/10.5281/zenodo.17761959}{https://doi.org/10.5281/zenodo.17761959} in the file 3d/3d-G-polytopes.txt. Nonetheless, we do not provide an analytical, rigorous proof that these are all the minimal $\mc{G}$ polytopes. 

Here are some comments on these results: The list of minimal $\mc{G}$ polytopes contain some examples on the boundary of the set we
explored,
so we believe it is probably not the complete set of minimal $\mc{G}$
polytopes. However, the number of minimal boxes decreases rapidly as the
first three vertices move further from the origin. Minimal polytopes
not in the $\mc{B}_6(p, q)$ style are a small subset (estimated at less
than 500, and much smaller than 4000, the number of $\mc{B}_6(p, q)$ 
$\mc{G}$ polytopes) of all the minimal $\mc{G}$ polytopes. So we
conclude that most of the minimal polytopes are of the $\mc{B}_6(p, q)$
form, and in the following sections we will mainly focus on such
examples. Moreover,  we find that minimal $\mc{G}$ polytopes
are concentrated in the pattern \be \{(1, 0, 0), (q-1, q, 0), (r, r,
q), (-a, -b, -c)\} \ee as the size of minimal $\mc{G}$ polytopes grows.

In total, we find 4454 inequivalent minimal $\mc{G}$ polytopes with 4 vertices, without proving completeness.

As for the minimal $\mc{G}$ polytopes with 5 vertices, we have
searched over $\mc{G}$ polytopes with 5 vertices and the maximal
elements of the first three vertices less than 7. There are 98 $\mc{G}$
polytopes in total that are minimal within this set, of which 78
$\mc{G}$ polytopes are of the form $\{(1,0,0),(0,1,0),(0,0,1),(0,-1,-p),(-1,0,-q)\}$ with $12\geq q\geq p\geq 1$. The other 20 minimal $\mc{G}$ polytopes are
listed in Table~\ref{t:minimal_v5}. Among them there
is a special kind of minimal $\mc{G}$ polytope with a similar pattern
\be
\{(1,0,0),(0,1,0),(0,n,n+1),(0,-1,-1),(-1,0,0)\}.
\ee
We further check these $\mc{G}$ polytopes with different $n$s. It
turns out that for $n<12$ we always get minimal $\mc{G}$ polytopes,
and for $n\geq13$ they are no longer minimal. If we believe the minimal
$\mc{G}$ polytopes are located relatively contiguously 
 in the parameter space of
elements of vertices, then we may 
 expect that we have found all the 98 minimal
$\mc{G}$ polytopes with exactly 5 vertices.
There may, of course, however, be  sporadic cases that arise outside
the region we have explored.

\begin{table}
\centering
\begin{tabular}{|c|}
\hline
Vertices\\
\hline
$\{(1,0,0),(0,1,0),(0,1,2),(0,-1,-1),(-1,0,0)\}$\\
$\{(1,0,0),(0,1,0),(1,1,2),(0,-1,0),(-3,-1,-2)\}$\\
$\{(1,0,0),(0,1,0),(1,1,2),(-1,-2,0),(-1,-3,-2)\}$\\
$\{(1,0,0),(0,1,0),(1,1,2),(0,-1,0),(-5,-1,-2)\}$\\
$\{(1,0,0),(0,1,0),(1,1,2),(-2,-1,0),(-5,-1,-2)\}$\\
$\{(1,0,0),(0,1,0),(1,1,2),(-1,-4,0),(-1,-5,-2)\}$\\
$\{(1,0,0),(0,1,0),(1,1,2),(0,-1,0),(-7,-1,-2)\}$\\
$\{(1,0,0),(0,1,0),(1,1,2),(-2,-1,0),(-7,-1,-2)\}$\\
$\{(1,0,0),(0,1,0),(1,1,2),(-4,-1,0),(-7,-1,-2)\}$\\
$\{(1,0,0),(0,1,0),(0,2,3),(0,-1,-1),(-1,0,0)\}$\\
$\{(1,0,0),(0,1,0),(1,1,3),(0,-1,0),(-2,-1,-3)\}$\\
$\{(1,0,0),(0,1,0),(0,3,4),(0,-1,-1),(-1,0,0)\}$\\
$\{(1,0,0),(0,1,0),(0,4,5),(0,-1,-1),(-1,0,0)\}$\\
$\{(1,0,0),(0,1,0),(0,5,6),(0,-1,-1),(-1,0,0)\}$\\
$\{(1,0,0),(0,1,0),(0,6,7),(0,-1,-1),(-1,0,0)\}$\\
$\{(1,0,0),(0,1,0),(0,7,8),(0,-1,-1),(-1,0,0)\}$\\
$\{(1,0,0),(0,1,0),(0,8,9),(0,-1,-1),(-1,0,0)\}$\\
$\{(1,0,0),(0,1,0),(0,9,10),(0,-1,-1),(-1,0,0)\}$\\
$\{(1,0,0),(0,1,0),(0,10,11),(0,-1,-1),(-1,0,0)\}$\\
$\{(1,0,0),(0,1,0),(0,11,12),(0,-1,-1),(-1,0,0)\}$\\
\hline
\end{tabular}
\caption{The list of minimal $\mc{G}$ polytopes with 5 vertices, and first three vertices different from
  $\{(1,0,0),(0,1,0),(0,0,1)\}$. Permutation redundancy has been
  removed. \label{t:minimal_v5}}
\end{table}

In total, we find $4454+98+1=4553$ minimal $\mc{G}$ polytopes corresponding to $4553$ maximal boxes.

\subsection{Monte Carlo in the largest box}
\label{sec:3d-largest}

In \cite{Wang:2020gmi}, the toric threefold base with the largest
$h^{1,1}(B_3)$, denoted by $\mc{B}_6{(84, 516)}$ in our notation, is
 analyzed; this base is
special in the whole landscape of 3d bases in F-theory, due to the
presence of an exceptionally 
 large number of different triangulations
(proved to be $>10^{45766}$ in \cite{Wang:2020gmi}). In the present
work, we ignore the triangulation details and treat the $\mc{B}_6{(84,
  516)}$ as a single base polytope. Hence it is natural to ask about
the total number of bases inside the box (\ref{maxver}), and compare
the number with the other 3d boxes. From the fact that this box corresponds to the base with the largest
 value of $h^{1,1}$, one might naively expect that it contains the
 most other bases; from the analysis of 2d bases in the previous
 section, however, we cannot rely on this, so in the next subsection
 we consider other bases as well.

There are $|R|=181203$ primitive rays in this base polytope. To generate compact base polytopes $\Delta_b$ in $\mc{B}=\mc{B}_6(84,516)$, we have to fix the following two points
\be
\{(-6,-6,1),(-6,37,-6)\}\,.
\ee
It is easy to check that the point $(-6,-6,1)$ has to be added into
the compact base polytope for the origin to remain in the interior, as this is the only point with a positive
$z$-coordinate. Similarly, if one removes the point $(-6,37,-6)$ from
the box $\mc{B}$, the convex hull of the rest of the points cannot contain
the origin $(0,0,0)$ in the interior.

After that, we randomly choose $(n-2)$ more rays to form a polytope with $n$ chosen points. Note that the $n$ points may not all be vertices. By repeating this process, we generate $N$ samples with $M$ of them giving a compact base polytope. 

In order to use (\ref{Ntot}) to calculate the estimated number of bases, it is important to plug in
\be
N_c=|R|^{n-2}\,,
\ee
and the $n,m$ in the permutation redundancy factor (\ref{red-factor}) should be replaced by $n-2,m-2$, because 2 vertices are preemptively fixed. We have
\be
\label{Ntot-fix2}
N_{\rm tot}(n)\approx\frac{|R|^{n-2}}{N}\cdot \left(\sum_{i=1}^M\frac{1}{R(l(\Delta_{b,i}),m-2,n-2)(1+R_{GL(3,\mb{Z})}(\Delta_{b,i}))}\right)\,.
\ee
The $GL(3,\mb{Z})$ redundancy factors will almost all equal to 0 in
any sampling run, because the sampled base polytopes  typically resemble the shape of the long and thin box $\mc{B}$.

\begin{figure}
\centering
\includegraphics[height=7cm]{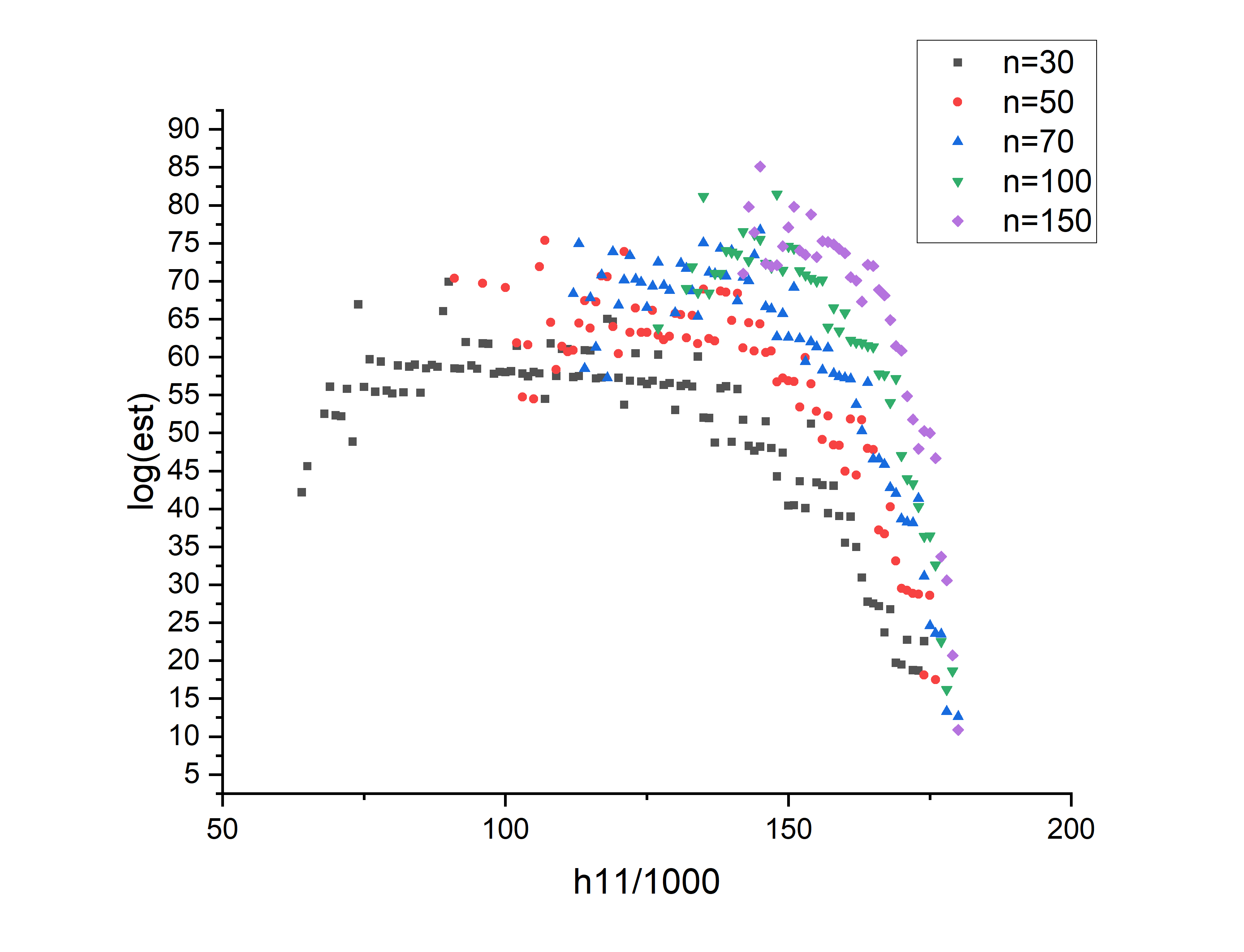}
\caption{The $\log_{10}$ of the estimated number of bases inside the largest box $\mc{B}_6(84,516)$, with two points $\{(-6,-6,1),(-6,37,-6)\}$ fixed, with respect to $h^{1,1}(B_3)$. The full data can be accessed on \href{https://doi.org/10.5281/zenodo.17761959}{https://doi.org/10.5281/zenodo.17761959} in the file MonteCarlo3d-large.7z. Different colors stand for data with different chosen $n$.}\label{f:large_fix2}
\end{figure}

We perform the Monte Carlo process with different numbers of sampled
points $n=\{30,50,70,100,150\}$, and plot the estimated number of
bases with respect to $h^{1,1}(B_3)$ in Figure~\ref{f:large_fix2}. The data can be found on \href{https://doi.org/10.5281/zenodo.17761959}{https://doi.org/10.5281/zenodo.17761959} in the file MonteCarlo3d-large.7z. One
can see that when $n$ increases, the sampled bases generally have a
larger $h^{1,1}(B_3)$, as the plotted curves move to the top
right. While for smaller $n$, the sampled bases have more variety and
occupy a wider range of $h^{1,1}(B_3)$.

\begin{figure}
\centering
\includegraphics[height=7cm]{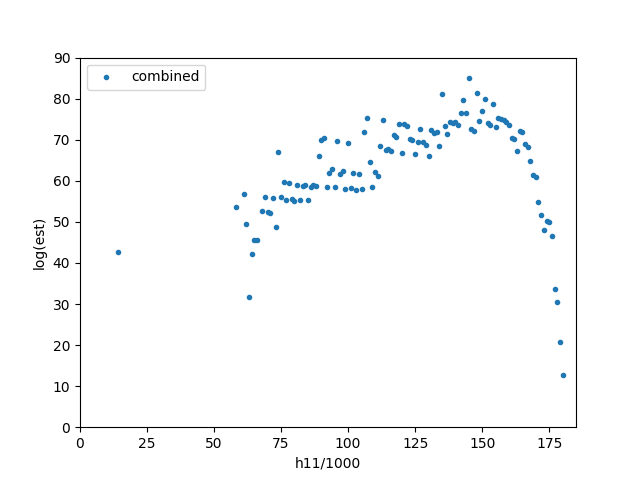}
\caption{The $\log_{10}$ of the estimated number of bases inside the largest box $\mc{B}_6(84,516)$, with two points $\{(-6,-6,1),(-6,37,-6)\}$ fixed, with respect to $h^{1,1}(B_3)$. One adds up cases for all different $n$s to obtain a full picture of how bases are distributed with respect to $h^{1,1}(B_3)$.}\label{f:combined_large}
\end{figure}

We can also combine the curves into a single
curve, shown in Figure~\ref{f:combined_large}. One can see in the figure that the peak
of the estimated number of bases lies at $h^{1,1}(B_3)\approx 1.4\times
10^{5}$, where $\sim 10^{85}$ different bases are 
 estimated to exist in the
largest box.

\begin{figure}
\centering
\includegraphics[height=7cm]{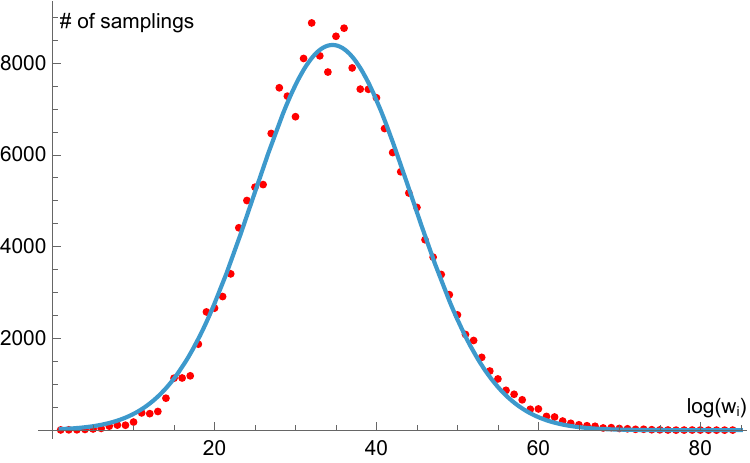}
\caption{The distribution of the weight factors $\log_{10}(w_i)$ in
  the Monte Carlo sampling of 3d base polytopes in the maximal box
  $\mc{B}_6(84,516)$, with the data in Figure~\ref{f:large_fix2}. The horizontal axis is $\log_{10}(w_i)$, while the vertical axis is the number of samplings in each bin. The blue curve is the optimal fitting with a Gaussian function.}\label{f:3d-L-weightfactor-dist}
\end{figure}

We verify the Gaussian distribution behaviours in Section~\ref{sec:Gaussian}. The distribution of weight factors across samplings, for the combined data of different $n$s, is shown in Figure~\ref{f:3d-L-weightfactor-dist}, with a comparison to the optimal Gaussian fitting curve.

\begin{figure}
\centering
\includegraphics[height=7cm]{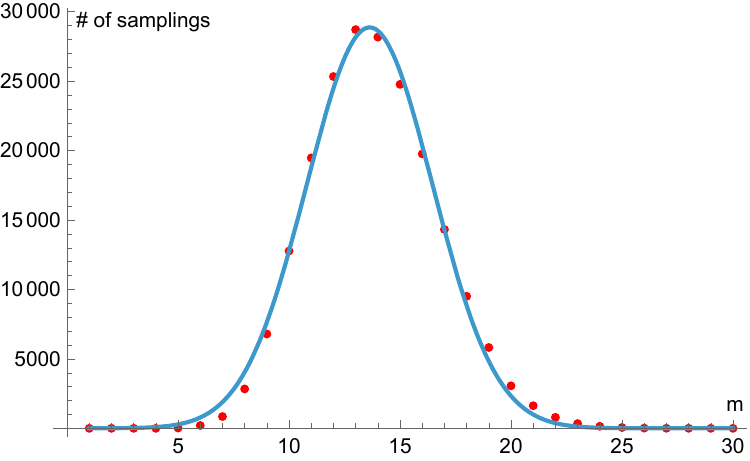}
\caption{The distribution of the number of vertices $m$ in the Monte Carlo sampling of 3d base polytopes in the maximal box $\mc{B}_6(84,516)$, with the data in Figure~\ref{f:large_fix2}. The horizontal axis is $m$, while the vertical axis is the number of samplings in each bin. The blue curve is the optimal fitting with a Gaussian function.}\label{f:3d-L-Nvertex-dist}
\end{figure}

We also verify the nearly Gaussian distribution of the number of samplings for the number of vertices $m$, in which we plot in Figure~\ref{f:3d-L-Nvertex-dist}.

\subsection{Monte Carlo in other boxes}
\label{sec:other-boxes}

In the largest box $\mc{B}_6(84,516)$, we mainly obtain bases with
$h^{1,1}(B_3)=1.1\times10^5$--$1.8\times10^5$. We now want to explore the regime of smaller bases. 

\begin{table}
\centering
\begin{tabular}{|c|l|c|}
\hline
Box &  Vertices other than $\{(-6,37,-6),(-6,-6,-6)\}$ & $|R|$\\
\hline
$\mc{B}_6(84, 516)$ & $\{(3606,-6, -6),(-6,-6, 1)\}$ & 181203\\
$\mc{B}_6(63, 387)$ & $\{(2706,-6-6),(-3,37,-6),(-3,-6,1),(-6,-6,1)\}$ & 136247\\
$\mc{B}_6(42, 258)$ & $\{(1806,-6,-6),(-6,31,-5),(0,37,-6),(0,-6,1),(-6,-6,1)\}$ & 91504\\
$\mc{B}_6(21, 129)$ & $\{(906,-6,-6),(3,37,-6),(-6,19,-3),(3,-6,1),(-6,-6,1)\}$ & 46665\\
\hline
\end{tabular}
\caption{Other examples of 3d $\mc{B}_6(p, q)$ base.}\label{t:others}
\end{table}

\begin{figure}
\centering
\includegraphics[height=7cm]{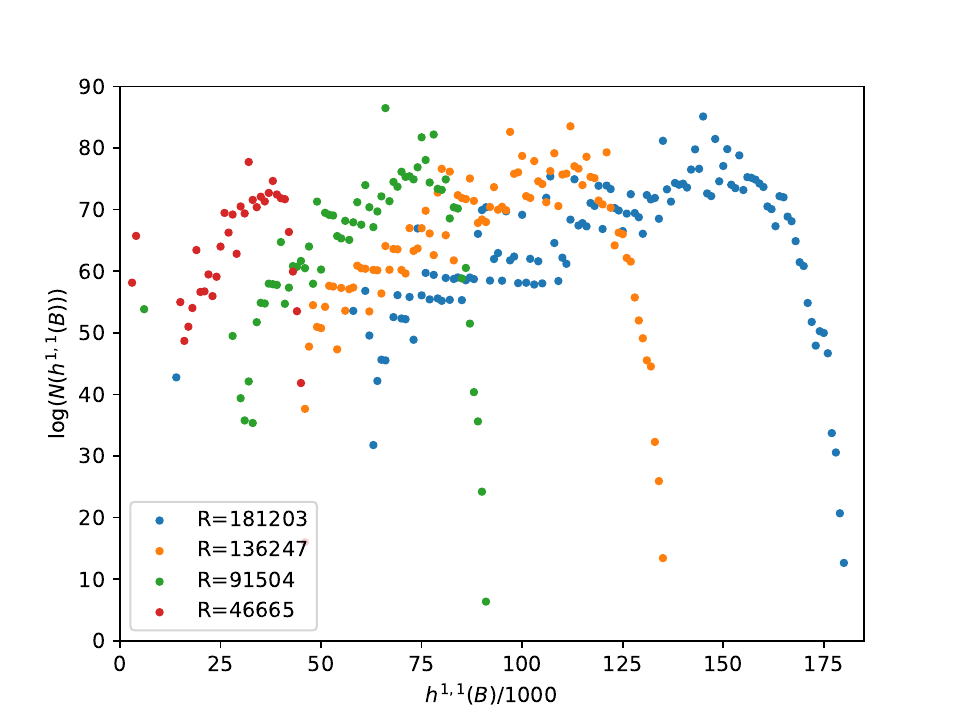}
\caption{The estimated number of $\log_{10}N_{\text{tot}}(h^{1,1}(B))$ in different boxes. The horizontal axis is divided in the unit of $10^3$.}\label{f:tsml}
\end{figure}

As shown in Figure~\ref{f:heat}, the largest box $\mc{B}_6(84, 516)$
lies at the bottom right of the figure. We consider the 1/4-, 1/2- and
3/4- positions of the diagonal line as examples of other boxes. The
detailed information of the boxes are listed in table
~\ref{t:others}. Similarly, we fix the first two vertices to be
$(-6,-6,1)\ ,\ (-6,37,-6)$ to ensure it is possible to get an allowed base polytope, and randomly choose the remaining $(n-2)$ vertices as in the
previous section. The $h^{1, 1}$ of different boxes ranges from
$30000$ to $180000$. Finally, we collect all data with different sizes
of boxes to form a more complete understanding of how bases with different
sizes and shapes are distributed. One can get the data on \href{https://doi.org/10.5281/zenodo.17761959}{https://doi.org/10.5281/zenodo.17761959} in the file MonteCarlo3d-medium.7z, MonteCarlo3d-small.7z and MonteCarlo-tiny.7z. The 
 distribution of estimated
number of bases with respect to $h^{1,1}$ from these four choices of
boxes
is shown in
Figure~\ref{f:tsml}.

We see that when sampling in different boxes, we are probing the
distribution of bases in certain distinct $h^{1,1}(B_3)$ regions. Distribution
patterns are similar in different boxes, and generally rise slowly as
$h^{1,1}(B_3)$ grows, reach a peak which is close to the size of
the box and then fall down rapidly when $h^{1,1}(B_3)$ is almost the size
of the whole box. We can also see from the data that the largest
estimated number of bases in different boxes are all around
$10^{85}$. Given that the total number of 3d maximal boxes is around
$\sim 10000$, it is reasonable to give an upper bound estimation of
the total number of 3d base polytopes as $\sim 10^{90}$.

If one tries the algorithm on boxes away from the rightmost diagonal
line, the distribution pattern is very similar to the examples along
 that line. We then conclude that given a box, we can explore
the landscape of polytopes with $h^{1, 1}$ a bit smaller than the
number of primitive rays inside the box (that is, $|R|$), and if we go
too far away from the peak, the estimation would be more rough due to
the low probability to find a small polytope in the box randomly,
which leads to larger uncertainty.

\subsection{Gauge group data}
\label{sec:gauge-group}

Most 3d toric bases formed from the polytopes we sample here have
multiple rigid (geometrically non-Higgsable) gauge factors \cite{Morrison:2014lca}
We are interested in the distribution of  rigid gauge groups that  are
useful in F-theory model building. For example, we want to figure out
how many $E_6$ group factors a polytope has on average.
For any given base polytope, the non-Higgsable gauge group can be
determined by computing the orders of vanishing of the Weierstrass
coefficients $f, g$ over each toric divisor (primitive ray in the
polytope), along with the monodromy conditions, using the methods
described in \cite{Taylor:2017yqr}.

Here we take
statistical data in the largest box for $n=10,20,30,50$. We treat this
data as a
single ensemble. We denote the set of all data with $h^{1,1}$ in the
interval $\left[1000h,1000(h+1)\right)$ as $\Delta(h)$. The average
  number of instances for a fixed gauge group factor $g$ in $\Delta(h)$ can be given as
  \be r_g(h)=\frac{\sum_{\Delta_i\in \Delta(h)}n(\Delta_i,g)\cdot
    w_i}{\sum_{\Delta_i\in \Delta(h)} w_i}.  \ee Here $n(\Delta_i,g)$
  is the number of group factors $g$ in polytope $\Delta_i$, and $w_i$ the
  weight factor of $\Delta_i$.

\begin{figure}
\centering
\includegraphics[height=7cm]{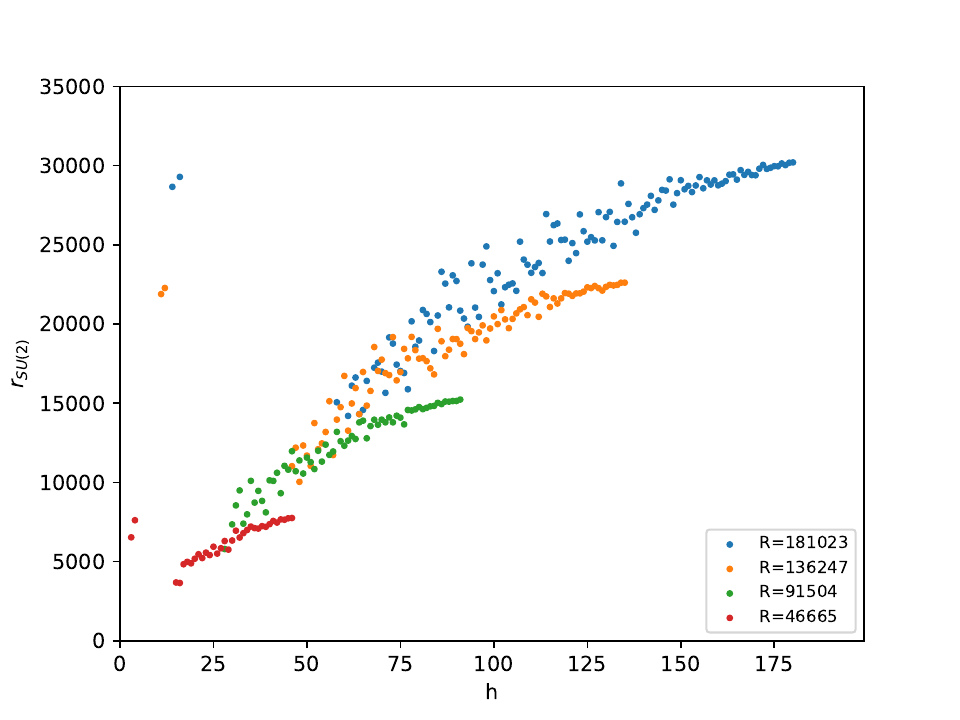}
\caption{The distribution of $r_{SU(2)}$ for $h$.}\label{f:su2}
\end{figure}

\begin{figure}
\centering
\includegraphics[height=7cm]{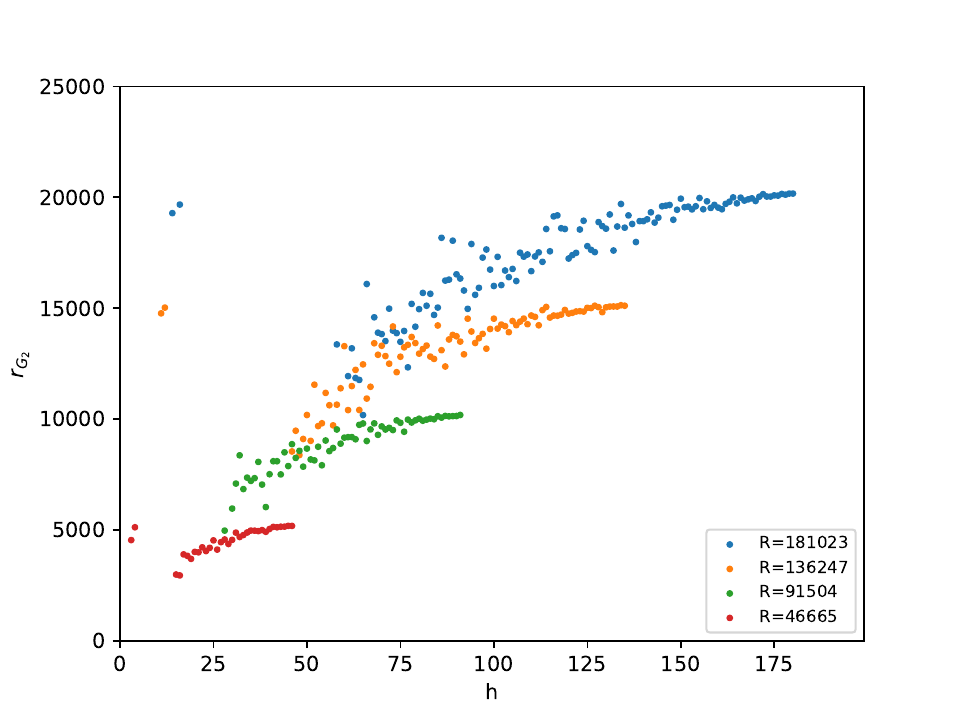}
\caption{The distribution of $r_{G_2}$ for $h$.}\label{f:g2}
\end{figure}

\begin{figure}
\centering
\includegraphics[height=7cm]{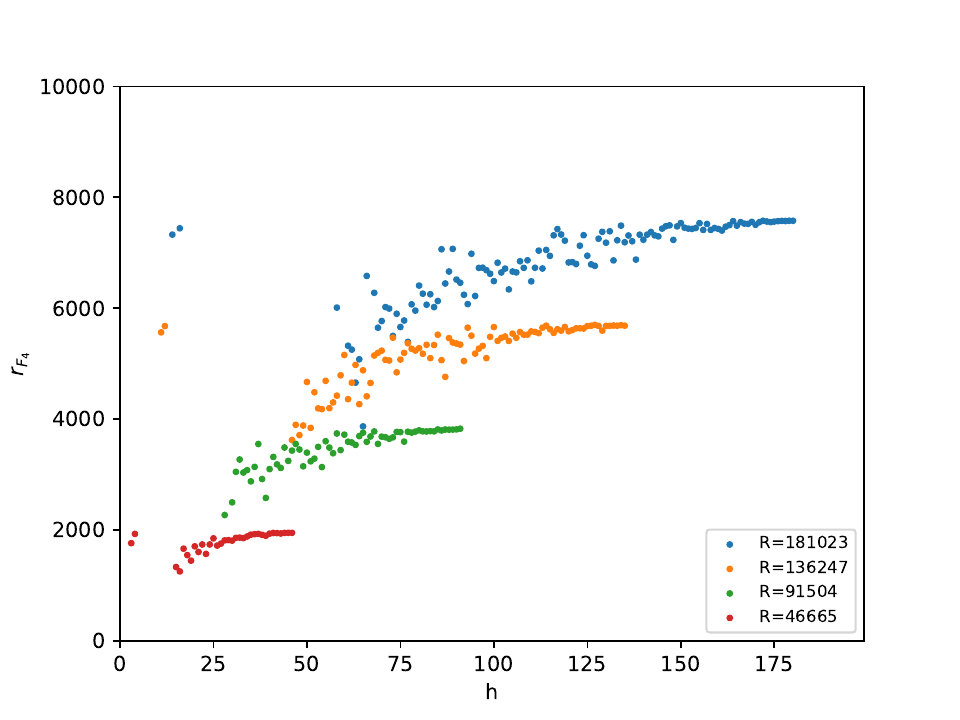}
\caption{The distribution of $r_{F_4}$ for $h$.}\label{f:f4}
\end{figure}

\begin{figure}
\centering
\includegraphics[height=7cm]{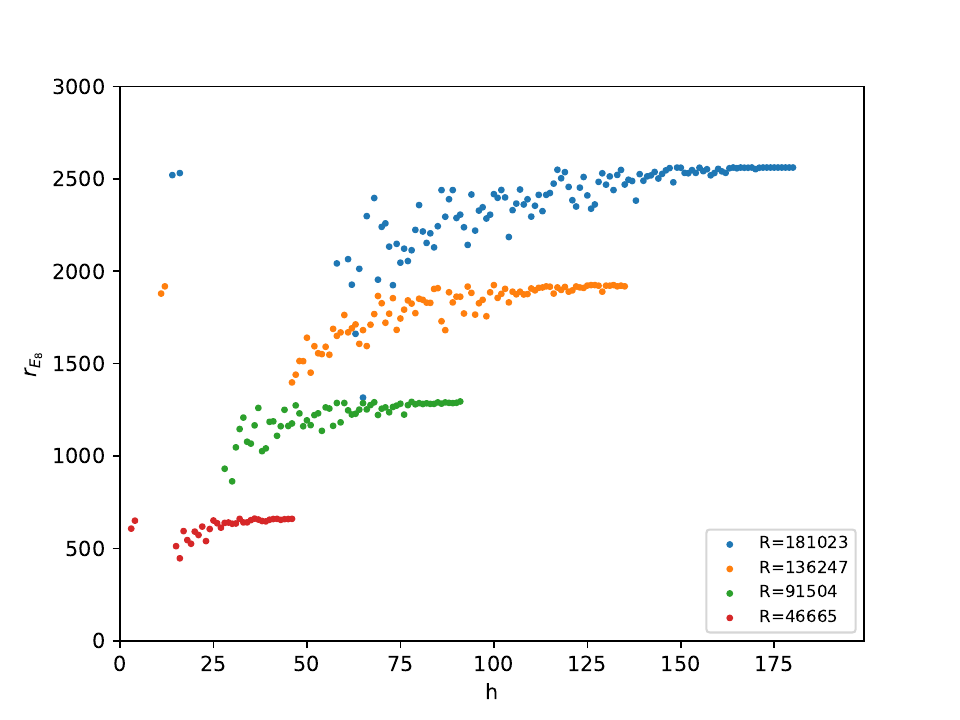}
\caption{The distribution of $r_{E_8}$ for $h$.}\label{f:e8}
\end{figure}

The different rigid gauge groups can be divided into two types. 
$SU(2), G_2, F_4,$ and $E_8$,  are the most common gauge groups for a
typical polytope. In Figures \ref{f:su2} to \ref{f:e8} we show the
dependence of the average number of these gauge group factors for a single base. The data is given in 3d\_gauge+weight files on \href{https://doi.org/10.5281/zenodo.17761959}{https://doi.org/10.5281/zenodo.17761959}. One
can clearly see that the  number of factors of these
gauge groups increases similarly as $h^{1, 1}$
grows, in a logarithm-like pattern. Notice that the common gauge groups
have different slopes. The typical ratio of $SU(2):G_2:F_4:E_8$ is
$50\%:33\%:13\%:4\%$. Note that the average numbers of 
 each gauge group factor are
different for different boxes even for the same $h$ value. This is due to our box
Monte-Carlo strategy, leading to typical bases with different shapes
and gauge groups in different boxes.

\begin{figure}
\centering
\includegraphics[height=7cm]{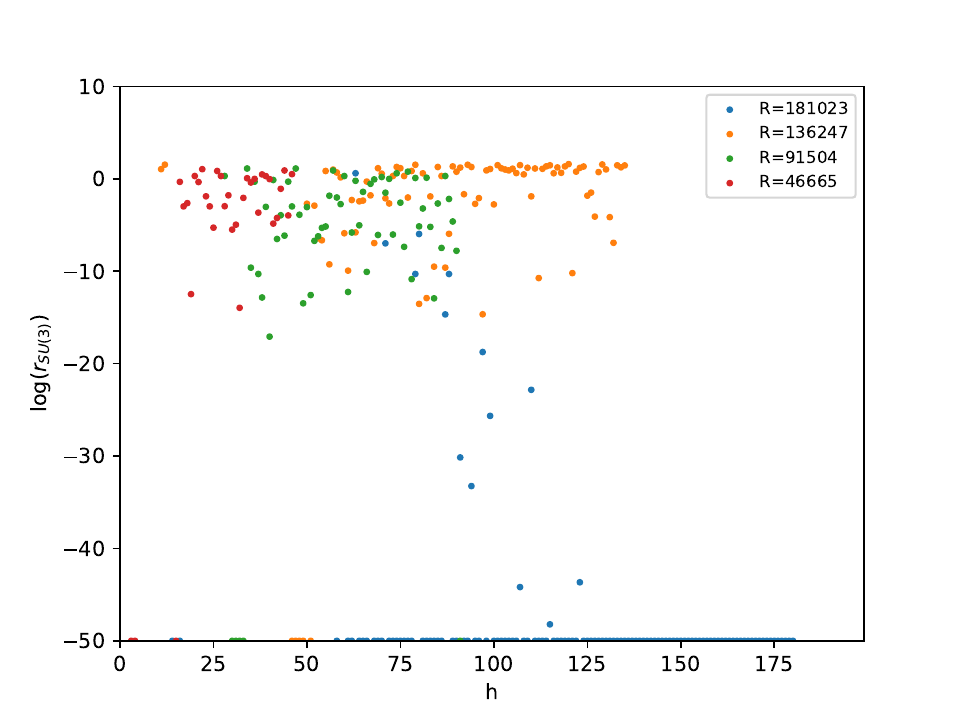}
\caption{The distribution of $\log r_{SU(3)}$ for $h$.}\label{f:su3}
\end{figure}

\begin{figure}
\centering
\includegraphics[height=7cm]{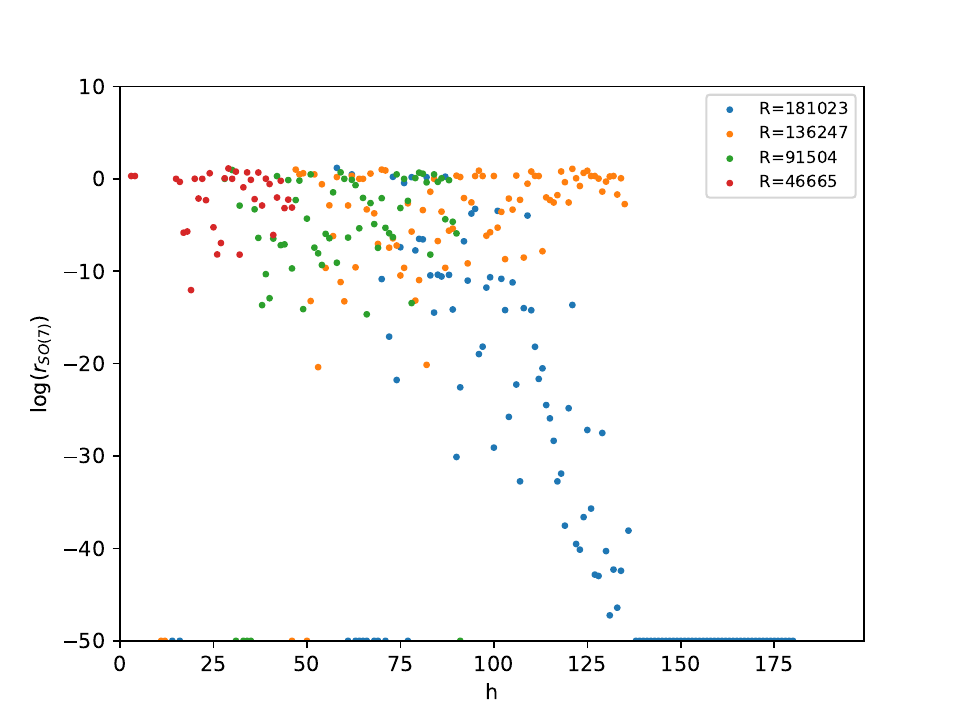}
\caption{The distribution of $\log r_{SO_7}$ for $h$.}\label{f:so7}
\end{figure}

\begin{figure}
\centering
\includegraphics[height=7cm]{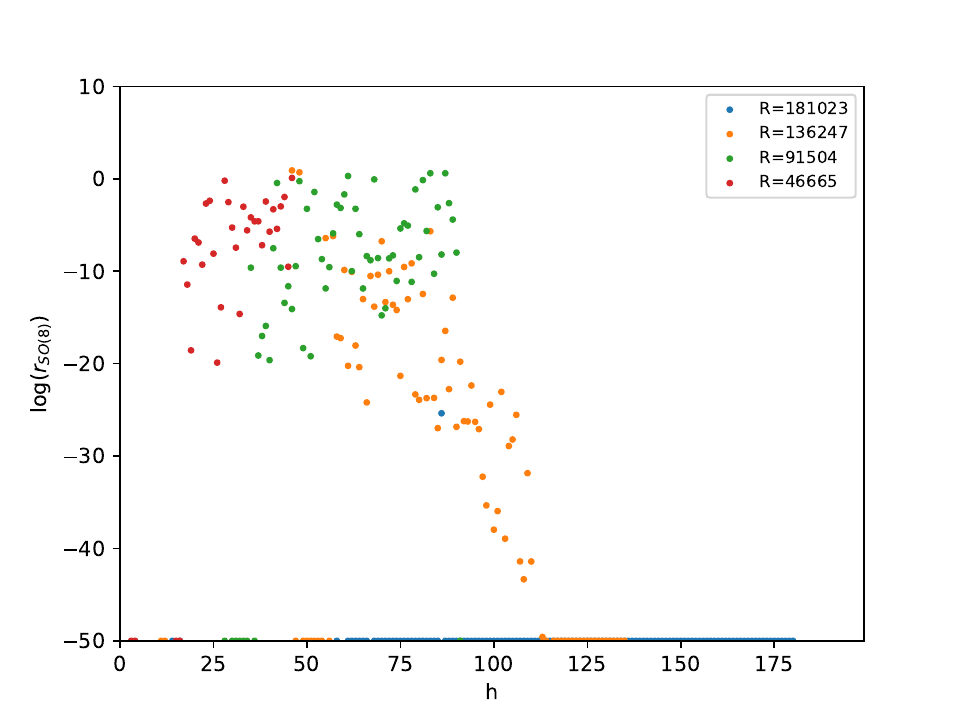}
\caption{The distribution of $\log r_{SO_8}$ for $h$.}\label{f:so8}
\end{figure}

\begin{figure}
\centering
\includegraphics[height=7cm]{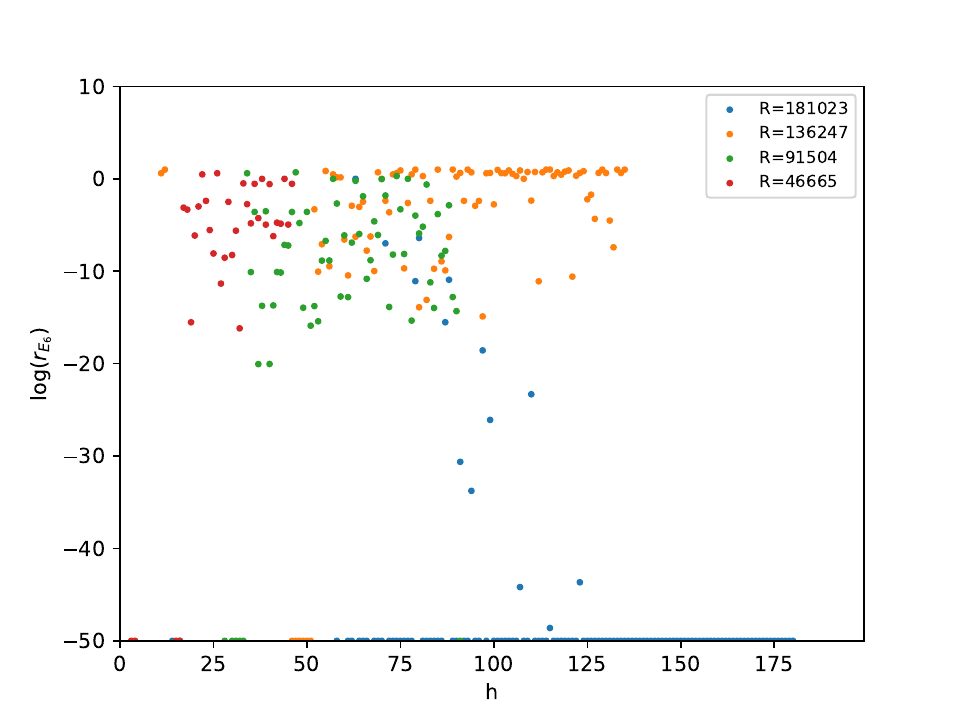}
\caption{The distribution of $\log r_{E_6}$ for $h$.}\label{f:e6}
\end{figure}

\begin{figure}
\centering
\includegraphics[height=7cm]{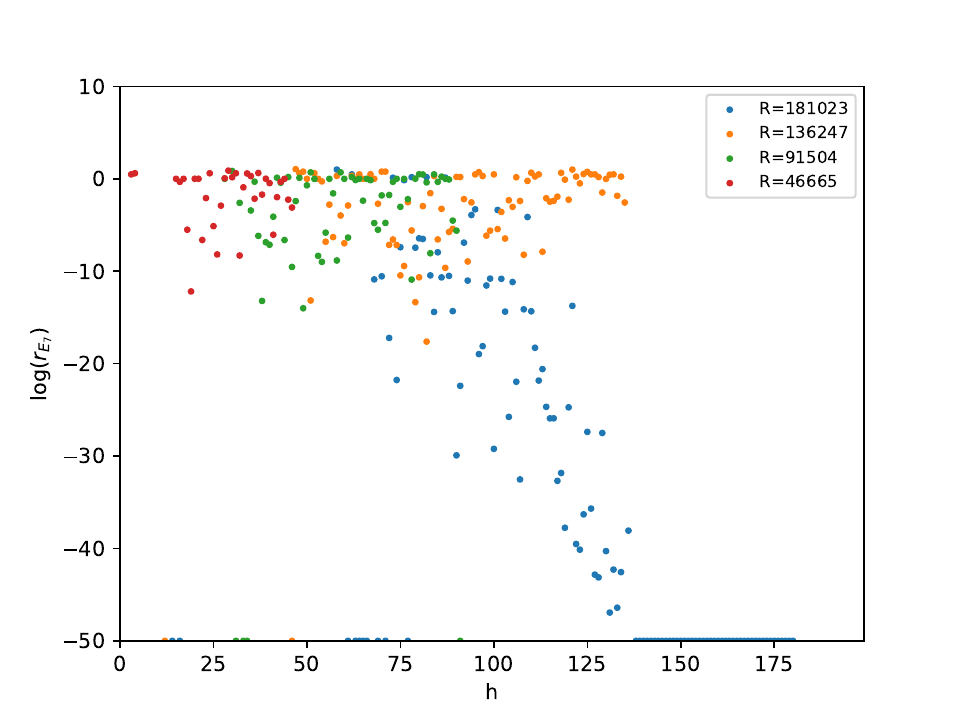}
\caption{The distribution of $\log r_{E_7}$ for $h$.}\label{f:e7}
\end{figure}

Besides the four common gauge groups, the remaining five gauge
groups $SU(3)$, $SO(7)$, $SO(8)$, $E_6$ and $E_7$ are relatively rare
on a base. We use a logarithmic diagram to show the average number of rare
gauge groups for different ranges of $h^{1, 1}$. Generally, the
average number of these gauge group factors per base is on the order of
$10^0$ or less. Another interesting fact is that the sampling result
from the largest box is qualitatively different from the smaller
boxes. The sampled base polytopes in the largest box have much smaller
numbers of rare gauge groups, and when $h^{1, 1} > 1.4\times 10^5$,
we find a complete absence of these rare gauge groups in our data! With the
results of the largest box, the rare gauge groups can be divided into
two types. For gauge groups $G = SO(7)$ and $ E_7$, even while the average
number of such gauge group factors is small and gradually falls as $h^{1,
  1}$ grows, we generally see nonzero distributions in
Figure~\ref{f:so7} and \ref{f:e7}. However, for $G=SU(3), SO(8), E_6$,
the nonzero distributions are more sparse, and in most cases they
would not appear in the bin of $h$; see Figures~\ref{f:su3},
\ref{f:so8} and \ref{f:e6}.

\subsection{Different base ensembles}
\label{sec:ensembles}

As in the 2d cases, we can also consider three different ensembles for
the compact 3d toric bases of 4d F-theory, labeled by ensemble (I),
(II), (III); we summarize and compare these ensembles here.
\begin{itemize}
\item (I) The set of smooth 3d bases without toric codimension-two
  (4,6) or codimension-three (8,12) loci, which was partially scanned
  in~\cite{Taylor:2015ppa,Taylor:2017yqr}, with many examples
  constructed explicitly in \cite{Halverson:2015jua,Halverson:2017ffz}. In counting the total
  number of these bases, the  numbers from these analyses
count different triangulations of the
  same 3d base polytope as different bases; this number explodes
  exponentially as $h^{1,1}(B_3)$ increases~\cite{Wang:2020gmi}, and
  can reach $>10^{45766}$ for $B_{\rm max}$. The total number of such
  bases after modding out different triangulations is unknown.

\item (II) The set of 3d base polytopes $\Delta_b$ in our ensemble. In
  the 3d case, unlike the 2d  case,
adding all primitive internal points as toric rays does
  not generally lead to a smooth base $B_3$. The generalized
  Hirzebruch threefold $\widetilde{\mb{F}}_2$ with rays $(1,0,0),
  (0,1,0), (0,0,1), (-1,-1,-2)$ is an example; as the 3d cone
  $\{(1,0,0), (0,1,0), (-1,-1,-2)\}$ has volume 2, there is a
  $\mb{C}^3/\mb{Z}_2$ orbifold singularity at the corresponding point
  on the toric variety.

Nonetheless, these bases still make a sensible ensemble, as they precisely correspond to the set of toric threefold bases with at worst terminal singularities ($\widetilde{\mb{F}}_2$ is a perfect example), since one cannot resolve it by adding more rays in the interior of $\Delta_b$. 

Similar to the 2d case, the bases in ensemble (II) also generally have toric codimension-two (4,6) and codimension-three (8,12) loci, that correspond to strongly coupled sectors.

From the Monte Carlo analysis we have described here, we estimate that there are $\sim 10^{90}$ base polytopes in the ensemble (II).

We should emphasize that this  number is not an estimate of the total number of 3d
bases with at worst terminal singularities (modding out flips and
flops), since (just as in 2d) it is possible to blow down $B_3$ without changing its
convex hull, and get other bases satisfying this condition. We should
rather interpret the set (II) we get as a set of starting points, with
a relatively homogeneous distribution, from which one can blow down to
get more different 3d bases for F-theory.

\item (III) The full set of the most general 3d bases where
  codimension-two (4,6) and codimension-three (8,12) loci are allowed,
  and the bases are not necessarily smooth. Similar to the 2d case,
  all such bases can be obtained by removing an arbitrary number of
  internal points in a 3d base polytope $\Delta_b$. For each
  $\Delta_b$ with number of primitive internal points $l(\Delta_b)$
  and $m$  vertices, one can construct $2^{l(\Delta_b)-m}$ bases from it. From the 3d Monte Carlo results, we estimate the total number of bases in the ensemble (III) to be $\sim 10^{54482}$, just from the largest base $B_{\rm max}$.

\end{itemize}

Without a good handle on the cosmological measure problem, we really do not know which of these ensembles provides a better  statistical measure on the F-theory landscape. Nonetheless, these provide different and useful windows on
the variety of structures that can appear in
this large landscape of consistent UV 4d quantum gravity theories, and there may be some utility to using each of these in studying various phenomenological aspects of the landscape (see, e.g., \cite{Cornell-axions}).
In the future a more physical measure could be computed via the quantum tunneling rate between different geometries, generalizing the weakly coupled IIB story in \cite{Gao:2025geq}, possibly in an eternal inflation setup~\cite{Hebecker:2020aqr}.

\section{Discussion}
\label{sec:discussions}

In this paper, we have developed a systematic Monte Carlo approach to
randomly sample $d$-dimensional lattice polytopes from a fixed set of
points. With the properly derived weight factors, one can study the
statistics of these lattice polytopes.
We applied this method to study toric 3d bases for 4d F-theory models.
Such statistical methods can also potentially
be applied to other combinatorics problems. For instance, we observe that the number of polytopes satisfying some general classes of conditions obey a Gaussian distribution with respect to the number of vertices $m$.  One can, for instance,
see this phenomenon in the Kreuzer-Skarke set of 4d reflexive polytopes.

A future direction would be applying the Monte Carlo
methods in this paper to generate reflexive polytopes of higher
dimensions. For instance, one can try to take the $k=1$ boxes and
sample $n$ points in them. If the resulting polytope is not reflexive,
one takes the $k=1$ polar dual and only keeps the lattice points in
that. The resulting lattice polytope would be a reflexive polytope by
definition, and all reflexive $d$-polytopes can be sampled in this
form.
 Additional filtering and redundancy factors should be included for
 such a statistical analysis of reflexive polytopes to work.
 Note, however, that the full $k = 1$ set of polytopes,
which includes Calabi-Yau hypersurface threefolds with various singularity types,
 may also be of
 direct physical interest \cite{AGNT}.

For $d=3$, we have first identified a set of ``maximal'' 3d boxes in which
one can sample the base polytopes. These boxes give rise to a set of
minimal dual $\mc{G}$-polytopes, which can also be used as starting points of
blow-ups in 3d toric minimal model program.
 While we did not construct the complete set
of maximal 3d boxes and corresponding minimal 3d $\mc{G}$-polytopes,
we believe that we have identified the majority of these boxes, and
the results from the boxes we analyzed should form a representative
sample; we leave a more complete analysis of the set of 3d boxes as a
challenge for future work --- this may be possible following an
approach similar to that used in \cite{Avram:1997rs} for the $k = 1$
case of maximal boxes.
Sampling the polytopes in the 
maximal 3d boxes gives a distribution on the set of toric threefold
bases for 4d F-theory models, in which only polytope data is
maintained, so that many bases that are birationally equivalent
(through, e.g., flips and flops),
with the same divisor and gauge group structures
are
identified.  This groups together bases with similar physics and
simplifies the combinatorial structure of this part of the F-theory
landscape, giving somewhere in the range of $10^{85}$--$10^{90}$
distinct base polytopes associated with 3d toric varieties that can support elliptic Calabi-Yau fourfolds.  On the other hand, 
this ensemble also broadens the class of toric bases somewhat beyond those considered previously.  In particular, note that that the base polytopes we found through this approach typically lead to bases with
terminal singularities on the base, such as the $U(3)$ orbifold singularity $\mb{C}^3/\mb{Z}_2(1,1,1)$~\cite{Morrison:2004fr}. The physical impact of these base
singularities in 4d F-theory  should be the subject of future
investigations. The base polytopes also possess non-minimal
codimension-two (4,6) loci in the elliptic fibrations over these
bases, which lead to strongly coupled F-theory sectors at
least at a high energy scale~\cite{Apruzzi:2018oge,Tian:2018icz}. It
would also be interesting to investigate the phenomenological impact of such
sectors.

As in previous studies of the F-theory landscape, one central takeaway
from this analysis is the ubiquitous presence of many rigid gauge
group factors in virtually all of the geometries sampled.  In addition
to providing natural gauge sectors for realizing the Standard Model,
most models have many additional ``hidden sector'' SU(2), $G_2, F_4$,
and $E_8$ gauge factors that could act as dark matter.  Furthermore,
the strongly coupled (4, 6) sectors that arise in the geometries
considered here should also be considered as potential dark matter
candidates.\footnote{We would like to thank Jean Du Plessis for
discussion on this point.}

In this new measure for sampling 3d bases, we  found in particular that the
presence of $E_6$ or $E_7$ gauge groups that can be useful in Standard
Model building \cite{Li:2021eyn,Li:2022aek,Li:2023dya} are not that uncommon, in particular if we are away
from the maximal box $\mc{B}_6(84,516)$; see Figure~\ref{f:e6} and
\ref{f:e7}.
For such Standard Model constructions, however, typically a non-toric base is
necessary, so for explicit model building one either needs  an alternative way to realize hypercharge
flux breaking and the Higgs sectors in order to actually construct
``generic'' MSSM on these toric base geometries, or to extend this analysis
to non-toric bases.  More generally, however, assuming that the distribution of gauge groups is not too different for more general (non-toric) base geometries, these statistics support the hypothesis that 
rigid $E_6$ and $E_7$ gauge groups are reasonably common in the 4d F-theory landscape, at least as features of polytope equivalence classes of bases.

As in previous studies, we find that of the rigid/non-Higgsable gauge
factors that contain the Standard Model, $E_8$ is by far the most
prevalent, even without accounting for flux multiplicities.  Thus,
despite the challenges posed by the presence of strongly coupled
matter at SCFT points \cite{Kim:2017toz,Apruzzi:2018oge,Tian:2018icz}, rigid $E_8$ factors remain the most
promising and abundant place to potentially find Standard Model-like physics in
the landscape of 4d F-theory models. 

While the set of non-toric threefold bases is
much less well understood than the set of toric bases,  the approach
of this paper, which groups together threefold
bases related by flops and flips, may be helpful as well in the non-toric context.

\acknowledgments
We would like to thank Jean Du Plessis, Shing Yan Li and Richard Nally for helpful discussions.
YNW and YHY are supported by National Natural Science
Foundation of China under Grant No. 12175004, No. 12422503 and by
Young Elite Scientists Sponsorship Program by CAST (2023QNRC001,
2024QNRC001).
The work of WT was supported by
 the U.S. Department of Energy, Office of Science, Office of High Energy Physics of U.S. Department of Energy under grant Contract Number  DE-SC0012567.
The work is supported by High-performance Computing Platform of Peking
University.
WT would like to thank the Santa Fe Institute, where some of this work was completed.
This work also made use of resources provided by subMIT
at the MIT physics Department \cite{bendavid2025submitphysicsanalysisfacility}.

\bibliographystyle{JHEP}
\bibliography{F-ref}

\end{document}